\documentstyle[epsfig]{mn2e}

\begin{document}

\title[WiggleZ survey: small-scale clustering]{The WiggleZ Dark Energy
  Survey: small-scale clustering of Lyman Break Galaxies at $z < 1$}

\author[Blake et al.]{\parbox[t]{\textwidth}{Chris
    Blake$^1$\footnotemark, Russell J.\ Jurek$^2$, Sarah Brough$^1$,
    Matthew Colless$^3$, Warrick Couch$^1$, Scott Croom$^4$, Tamara
    Davis$^{2,5}$, Michael J.\ Drinkwater$^2$, Duncan Forbes$^1$, Karl
    Glazebrook$^1$, Barry Madore$^6$, Chris Martin$^7$, Kevin
    Pimbblet$^2$, Gregory B.\ Poole$^1$, Michael Pracy$^{1,8}$, Rob
    Sharp$^3$, Todd Small$^7$ and David Woods$^{9,10}$} \\ \\ $^1$
  Centre for Astrophysics \& Supercomputing, Swinburne University of
  Technology, P.O. Box 218, Hawthorn, VIC 3122, Australia \\ $^2$
  Department of Physics, University of Queensland, Brisbane, QLD 4072,
  Australia \\ $^3$ Anglo-Australian Observatory, P.O. Box 296,
  Epping, NSW 2121, Australia \\ $^4$ School of Physics, University of
  Sydney, NSW 2006, Australia \\ $^5$ Dark Cosmology Centre, Niels
  Bohr Institute, University of Copenhagen, Juliane Maries Vej 30,
  DK-2100 Copenhagen, Denmark \\ $^6$ Observatories of the Carnegie
  Institute of Washington, 813 Santa Barbara St., Pasadena, CA 91101,
  United States \\ $^7$ California Institute of Technology, MC 405-47,
  1200 East California Boulevard, Pasadena, CA 91125, United States \\
  $^8$ Research School of Astronomy and Astrophysics, Australian
  National University, Weston Creek, ACT 2600, Australia \\ $^9$
  School of Physics, University of New South Wales, Sydney, NSW 2052,
  Australia \\ $^{10}$ Department of Physics \& Astronomy, University
  of British Columbia, 6224 Agricultural Road, Vancouver, B.C., V6T
  1Z1, Canada}

\maketitle

\begin{abstract}
  The WiggleZ Dark Energy Survey is a large-scale structure survey of
  intermediate-redshift UV-selected emission-line galaxies scheduled
  to cover 1000 deg$^2$, spanning a broad redshift range $0.2 < z <
  1.0$.  The main scientific goal of the survey is the measurement of
  baryon acoustic oscillations (BAO) in the galaxy clustering pattern
  at a significantly higher redshift than previous studies.  The BAO
  may be applied as a standard cosmological ruler to constrain dark
  energy models.  Based on the first $20\%$ of the dataset, we present
  initial results concerning the small-scale clustering of the WiggleZ
  targets, together with survey forecasts.  The WiggleZ galaxy
  population possesses a clustering length $r_0 = 4.40 \pm 0.12 \,
  h^{-1}$ Mpc, which is significantly larger than $z=0$ UV-selected
  samples, with a slope $\gamma = 1.92 \pm 0.08$.  This clustering
  length is comparable to $z=3$ Lyman Break Galaxies with similar UV
  luminosities.  The clustering strength of the sample increases with
  optical luminosity, UV luminosity and reddening rest-frame colour.
  The full survey, scheduled for completion in 2010, will map an
  effective volume $V_{\rm eff} \approx 1$ Gpc$^3$ (evaluated at a
  scale $k = 0.15 \, h$ Mpc$^{-1}$) and will measure the
  angular-diameter distance and Hubble expansion rates in three
  redshift bins with accuracies $\approx 5\%$.  We will determine the
  value of a constant dark energy equation-of-state parameter, $w_{\rm
    cons}$, with a higher precision than existing supernovae
  observations using an entirely independent technique.  The WiggleZ
  and supernovae measurements lie in highly complementary directions
  in the plane of $w_{\rm cons}$ and the matter density $\Omega_{\rm
    m}$.  The forecast using the full combination of WiggleZ,
  supernovae and CMB datasets is a marginalized error $\Delta w_{\rm
    cons} = 0.07$, providing a robust and precise measurement of the
  properties of dark energy including cross-checking of systematic
  errors.
\end{abstract}

\section{Introduction}
\renewcommand{\thefootnote}{\fnsymbol{footnote}}
\setcounter{footnote}{1}
\footnotetext{E-mail: cblake@astro.swin.edu.au}

The large-scale structure of the Universe is one of the pillars of our
modern understanding of cosmology, encoding information about the
contents and evolution of the Universe, the physics of the growth of
density fluctuations with time, and the formation and evolution of
galaxies within the underlying network of dark matter haloes.  In
particular, the large-scale clustering pattern of galaxies is
sensitive to the properties of the cosmic dark energy component which
is currently poorly understood.  Dark energy influences both the rate
of growth of structure and the geometrical distance-redshift
relations.  One of the cleanest probes of dark energy is to delineate
as a function of redshift the apparent tangential and radial size of
the baryon acoustic oscillation scale, a known ``standard ruler''
preferred separation imprinted into the galaxy distribution (Cooray et
al.\ 2001; Eisenstein 2002; Blake \& Glazebrook 2003; Seo \&
Eisenstein 2003; Hu \& Haiman 2003; Linder 2003; Glazebrook \& Blake
2005).  This cosmological probe is helping to motivate a new
generation of massive spectroscopic galaxy surveys.

Cosmic structure has been mapped out by a succession of galaxy
redshift surveys of increasing size and depth.  The local Universe
(redshifts $z < 0.2$) has been studied in exquisite detail by the
2-degree Galaxy Redshift Survey (2dFGRS; Colless et al.\ 2001) and the
Sloan Digital Sky Survey (SDSS; York et al.\ 2000).  The SDSS Luminous
Red Galaxy component extended this programme to a mean redshift $z
\approx 0.35$ using a specific type of tracer galaxy (Eisenstein et
al.\ 2001).  Indeed, the cosmological conclusions reached should be
independent of the galaxy type used, given that the ``bias'' with
which galaxies trace the underlying dark matter fluctuations is
expected to be a simple linear function on large scales (Coles 1993;
Scherrer \& Weinberg 1998).  In this sense, the choice of the ``tracer
population'' of galaxies can be determined by observational
considerations such as telescope exposure times, the availability of
input imaging data for target selection, and secondary science goals.

The WiggleZ Dark Energy Survey, using the AAOmega multi-object
spectrograph at the 3.9m Anglo-Australian Telescope (AAT), is designed
as the next leap forwards in redshift coverage, targetting the range
$0.2 < z < 1.0$.  The survey is scheduled to cover a sky area of 1000
deg$^2$, mapping a cosmic volume $V \sim 1$ Gpc$^3$ sufficient to
measure the imprint of baryon oscillations in the clustering pattern
at a significantly higher redshift than has been previously achieved
by 2dFGRS (Cole et al.\ 2005; Percival et al.\ 2007) and SDSS
(Eisenstein et al,\ 2005; Huetsi 2006; Percival et al.\ 2007;
Gaztanaga et al.\ 2008).  The survey redshift range is motivated by
the optimal redshift location for testing a cosmological constant
model in a spatially-flat Universe (Parkinson et al.\ 2007), which is
the sensible initial hypothesis to reject in the dark energy parameter
space.  The target galaxy population is bright emission-line galaxies
selected from UV imaging by the Galaxy Evolution Explorer (GALEX)
satellite (Martin et al.\ 2005).  This choice is motivated by the
short (1-hr) exposure times required to obtain redshifts at the AAT.
The survey commenced in August 2006 and is scheduled to finish in July
2010, using the equivalent of 165 clear nights of telescope time (220
awarded nights).  Secondary science goals involve the study of star
formation and galaxy evolution as a function of redshift and
environment.

In this initial study we focus on the small-scale clustering
properties of the first $20\%$ of the WiggleZ sample.  The clustering
strength is an important parameter in the survey design and
cosmological parameter forecasts: the signal-to-noise with which we
can recover the galaxy power spectrum depends on the bias of the
galaxies with respect to the dark matter fluctuations, which affects
the balance between sample variance and shot noise in the power
spectrum error budget.  These initial clustering measurements allow us
to determine the bias parameter and complete the survey forecast.

Furthermore, the joint UV-optical selection in the redshift interval
$0.2 < z < 1$ places the WiggleZ survey in an interesting location in
the parameter space of galaxy evolution.  In this context, the
clustering strength of a set of galaxies provides a direct indication
of the density of the environment or (equivalently) the typical mass
of the dark matter haloes hosting the galaxies.  The clustering
strength of UV-selected samples has been studied at low redshift $z
\approx 0$ (Milliard et al.\ 2007; Heinis et al.\ 2007) and the
corresponding rest-frame samples have been selected at much higher
redshift $z \approx 3$ through studies of the clustering of Lyman
Break Galaxies (LBGs; e.g.\ Giavalisco \& Dickinson 2001; Ouchi et
al.\ 2001; Arnouts et al.\ 2002; Foucaud et al.\ 2003; Adelberger et
al.\ 2005; Allen et al.\ 2005; Ouchi et al.\ 2005; Lee et al.\ 2006;
Yoshida et al.\ 2008).  The WiggleZ survey samples a redshift range
which is intermediate to these previous studies.  Moreover, the
clustering strength of optically-selected star-forming galaxies at
high redshift has been studied over small areas by the Deep
Extragalactic Evolutionary Probe (DEEP2) project (Coil et al.\ 2008)
and the VIMOS VLT Deep Survey (VVDS; Meneux et al.\ 2006).  WiggleZ is
mapping an area $\sim 100$ times larger, and is therefore able to
measure accurately the clustering strength of the most luminous
star-forming galaxies, for which these smaller surveys are limited by
small-number statistics and sample variance.

The backdrop to these studies is the recent concept of ``down-sizing''
(Cowie et al.\ 1996; Glazebrook et al.\ 2004; van Dokkum et al.\ 2004)
whereby the stars in more massive galaxies appear to have formed
earlier, and the typical mass of the most actively star-forming
galaxies is expected to decrease with time.  A recent study of LBGs
(Yoshida et al.\ 2008) has emphasized the importance of studying the
clustering segregration with both UV and optical luminosities, which
crudely trace ongoing star formation rate and stellar mass,
respectively.  WiggleZ is well-suited for undertaking such studies
over the redshift range $0.2 < z < 1.0$.

\begin{figure*}
\center
\epsfig{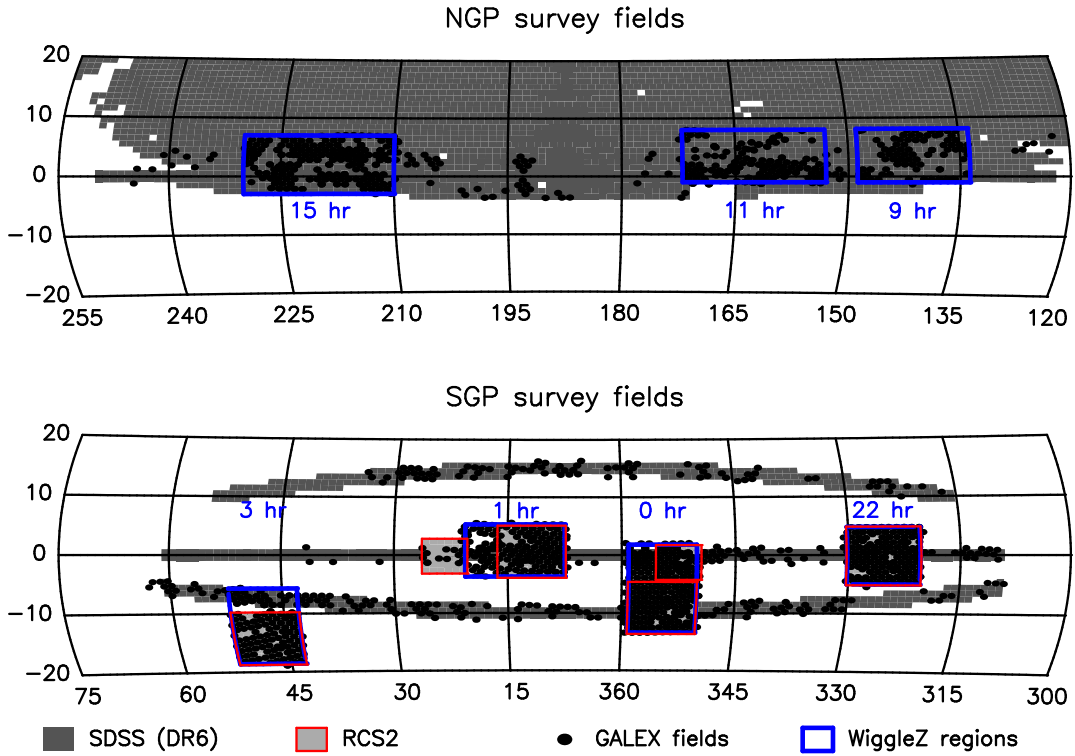}
\caption{The sky distribution of the seven WiggleZ survey regions
  compared to the coverage of the SDSS, RCS2 and GALEX Medium Imaging
  Survey at the end of 2008.}
\label{figsurvey}
\end{figure*}

The plan of this paper is as follows: in Section \ref{secdata} we
introduce the WiggleZ survey strategy and target selection and
describe the data sample used in this study.  In Section
\ref{secanalysis} we describe the methodology used to produce the
small-scale clustering measurement including the generation of random
(unclustered) realizations of the dataset with the correct selection
functions and redshift completeness map.  We also explain how we
derive the statistical error in the clustering measurement.  We
present the clustering results in Section \ref{secresults} (split by
redshift, absolute magnitude and rest-frame colour) together with
initial comparisons to other studies.  Section \ref{secforecast}
contains the cosmological parameter forecasts for the full WiggleZ
survey, and Section \ref{secconc} summarizes our conclusions.  When
converting redshifts to co-moving co-ordinates we assume a
spatially-flat Universe with cosmological parameters $\Omega_{\rm m} =
0.3$ and $\Omega_\Lambda = 0.7$.

\section{Data}
\label{secdata}

The design and implementation of the WiggleZ Dark Energy Survey will
be fully described in a forthcoming ``survey paper'' (Drinkwater et
al.\ 2009, in preparation) which will accompany our mid-term Data
Release.  We include a brief outline here both for ease of reference
and to emphasize the key points relevant to the small-scale clustering
analysis.

\subsection{WiggleZ survey strategy}

The WiggleZ survey strategy is to harvest low signal-to-noise spectra
of a large number of UV-selected emission-line galaxies in relatively
short exposure times (1-hr integrations at the AAT).  The survey
tolerates a relatively low redshift completeness of $70\%$ but
generates a large statistical sample of galaxy redshifts.  The survey
goal is to cover $1{,}000$ deg$^2$ of the equatorial sky, gathering
$\sim 350{,}000$ spectra of which $\sim 245{,}000$ are expected to
yield successful redshifts.  The survey was designed such that the
average galaxy number density $n$ is related to the amplitude of the
galaxy clustering power spectrum $P_{\rm gal}$ on the relevant baryon
oscillation scales by $n \sim 1/P_{\rm gal}$, implying that the
contributions of sample variance and shot noise to the clustering
error are equal.  This is the optimal survey strategy for fixed number
of galaxies.

The WiggleZ survey area, illustrated in Figure \ref{figsurvey}, is
split into seven equatorial regions to facilitate year-round
observing.  We require that each region should possess a minimum
angular dimension of $\sim 10$ deg, corresponding to a spatial
co-moving scale that exceeds by at least a factor of two the standard
ruler preferred scale [which projects to (8.5, 4.6, 3.2, 2.6) deg at
$z = (0.25, 0.5, 0.75, 1.0)$].  The survey coverage within individual
regions should also be highly ($> 70\%$) contiguous, otherwise the
significance of the detection of the acoustic features is degraded by
convolution with the survey window function.  The survey duration is
forecast to be $\sim 165$ clear nights between August 2006 and July
2010, using the multi-object capability of the 2dF positioner system
coupled to the AAOmega spectrographs (Saunders et al.\ 2004; Sharp et
al.\ 2006).

Galaxy redshifts are obtained from the bright emission lines
associated with star-forming galaxies, in particular redshifted [OII]
3727\AA, H$\beta$ 4861\AA\, and [OIII] 4959\AA, 5007\AA.
Low-resolution (5\AA\ FWHM) spectra are obtained spanning the
(observed-frame) wavelength range $5500 - 9500$\AA, hence the majority
of successful redshifts in the range $z < 0.95$ are confirmed by
multiple emission lines.  Single-line redshifts are almost invariably
[OII], for which we usually resolve the doublet in the range $z >
0.8$, increasing our confidence in the line identification.  Redshifts
are obtained by visual inspection of each spectrum using the
interactive software tool ``runz'', and are classified by a quality
flag $1 \le Q \le 5$, where the range $Q \ge 3$ denotes a ``reliable''
redshift (see Colless et al.\ 2001).  The fraction of stellar
contamination is very small ($< 1\%$) and we find a similarly low
fraction of high-redshift quasar interlopers.  The galaxy continuum is
typically detected with low signal-to-noise (an average of $S/N \sim
1$ per resolution element).

\subsection{WiggleZ survey target selection}
\label{sectarsel}

WiggleZ targets are chosen by a joint UV-optical selection.  The
primary selection dataset is the Medium Imaging Survey undertaken by
the Galaxy Evolution Explorer (GALEX) UV satellite, which provides
typical exposure times of 1500 seconds in two filter bands, $FUV$
($1350 - 1750$\AA) and $NUV$ ($1750 - 2750$\AA).  The GALEX
point-spread function is too broad to allow for accurate placement of
the spectrograph optical fibres, therefore the UV imaging is
cross-matched with optical data.  For our NGP regions, the SDSS
imaging data are used.  For our SGP regions, the SDSS $2.5^\circ$
stripes are too narrow compared to the preferred baryon oscillation
scale hence we use imaging data from the second Red Cluster Sequence
(RCS2) project instead (Yee et al.\ 2007).  Sources are cross-matched
between the GALEX and optical catalogues with a matching tolerance of
2.5 arcsec (which produces a negligible fraction of incorrect
matches).  In each imaging dataset, the majority of galaxies possess
relatively low signal-to-noise ($S/N = 3 - 5$) but their detection in
both datasets ensures a robust sample.  We note that acquisition of
our eventual requirement of $\sim 1250$ GALEX orbits of data is still
ongoing.  About $70\%$ of this total had been obtained at the end of
2008.

Targets are chosen from the UV-optical matched sample using a series
of magnitude and colour cuts.  These cuts are tuned to optimize the
fraction of targets lying at high redshift $z > 0.5$.  Firstly the
galaxy magnitudes are de-reddened using standard dust corrections
based on the local value of $E(B-V)$ measured from the Schlegel,
Finkbeiner \& Davis (1998) dust maps.  The primary GALEX selection
criterion is a red $FUV - NUV$ colour ($FUV - NUV > 1$ or $FUV$
drop-out), motivated by the Lyman Break passing through the $FUV$
filter for $z > 0.5$, and tuned by looking at galaxy templates.  At
the depth of the Medium Imaging Survey this colour is noisy, resulting
in a significant amount of contamination by low-redshift ($z < 0.5$)
galaxies which are partially removed by the additional cuts described
below.  We also impose a faint UV magnitude limit $NUV < 22.8$ and an
additional signal-to-noise requirement $S/N > 3$ for the detected NUV
flux (which becomes relevant for fields with unusually high dust
content or low exposure time).  The GALEX field-of-view is circular
with radius $\sim 0.6$ deg; we only select sources within the central
$0.55$ deg because of concerns over the photometry at the edge of the
field.

Our primary optical selection cuts are derived from SDSS $r$-band
imaging.  We require a UV-optical colour in the range $-0.5 < NUV - r
< 2$ based on the expected model tracks of star-forming galaxies.  We
impose a bright $r$-band limit $20 < r < 22.5$; the UV-optical colour
cut implies that the median optical magnitude of our targets is $r
\sim 21.5$.  Finally we increase the high-redshift efficiency by
introducing optical colour cuts.  Different cuts are used for the SDSS
and RCS2 regions, governed by the available imaging bands and depths.
For the SDSS regions analyzed in this paper, we apply cuts for those
(brighter) galaxies with good detections in the SDSS $g$- and
$i$-bands.  Specifically, for targets with $g < 22.5$ and $i < 21.5$
we reject galaxies in the colour space defined by $r - i < g - r -
0.1$ and $r - i < 0.4$ which is occupied by low-redshift galaxies both
theoretically and empirically (more details will be given in
Drinkwater et al.\ 2009, in preparation).  The final fraction of $z >
0.5$ galaxies obtained is $\approx 70\%$.  The redshift distribution
is displayed in Figure \ref{fignz}.

An average of 3-4 pointings of the 2dF spectrograph per patch of sky
is required in order to achieve the required target density of 350
deg$^{-2}$.  For any observing run the optimal placement of field
centres (based on the current availability of targets) is achieved
using the ``Metropolis'' (simulated annealing) algorithm (Campbell,
Saunders \& Colless 2004).  Galaxies are prioritized for spectroscopic
follow-up on the basis of optical $r$-band magnitude, in the sense
that fainter targets are observed first.  The motivation for this
strategy is to combat the potential inefficiency of ``mopping up''
residual galaxies in the final pointing for any patch of sky: the
brighter remaining galaxies can be observed in a shorter exposure time
by configuring fewer fibres.

\subsection{WiggleZ July 2008 data sample}

In this paper we analyze the subset of the WiggleZ sample assembled
from our first observations in August 2006 up until the end of the 08A
semester (July 2008).  At this point we had utilized 108 of our
allocated nights, of which the equivalent of 70 nights were clear.
The available galaxy database included $\approx 97{,}000$ reliable ($Q
\ge 3$) WiggleZ unique galaxy redshifts.

In this analysis we only use those galaxies lying in the SDSS regions
of our optical imaging because work is still on-going on the RCS2
portion of the angular selection function.  Specifically, we include
the WiggleZ 9-hr, 11-hr, 15-hr and 0-hr (SDSS) regions illustrated in
Figure \ref{figsurvey}.  The number of existing AAOmega pointings in
these regions is (42, 98, 140, 48).  The numbers of galaxies in each
region with reliable redshifts satisfying the final survey selection
criteria are (5782, 14873, 21629, 4383), constituting a total sample
of $N = 46{,}667$ for this initial analysis (about $20\%$ of the final
sample).

\section{Analysis}
\label{secanalysis}

\subsection{Correlation function estimator}
\label{seccorrest}

We quantify the small-scale clustering of the galaxy distribution
using a standard set of techniques based on the 2-point correlation
function.  This statistic compares the number of observed close galaxy
pairs to that expected by random chance, as a function of spatial
separation.  The key requirement is an ensemble of random
(unclustered) realizations of the survey possessing the same selection
function as the observed galaxy distribution.  With this in place we
convert the data (D) and random (R) galaxy angle-redshift
distributions into a grid of co-moving co-ordinates $(x,y,z)$ using an
assumed cosmological model (we use a flat model with $\Omega_{\rm m} =
0.3$).  We then bin the number of data-data (DD), data-random (DR) and
random-random (RR) pairs in the two-dimensional space of separation
perpendicular to the line-of-sight (denoted by $\sigma$) and parallel
to the line-of-sight (denoted by $\pi$).  This decomposition is
motivated by the influence of galaxy peculiar velocities
(redshift-space distortions) which shift galaxies in $\pi$, but not in
$\sigma$.  Each of our random realizations contains the same number of
targets as the data sample, and is generated by a method described
below.  The pair counts $DR$ and $RR$ are determined by averaging over
10 random realizations.

The 2D redshift-space correlation function $\xi_z(\sigma,\pi)$ is
derived using the estimator proposed by Landy \& Szalay (1993):
\begin{equation}
  \xi_z(\sigma,\pi) = \frac{DD(\sigma,\pi) - 2 \, DR(\sigma,\pi) +
    RR(\sigma,\pi)}{RR(\sigma,\pi)}
\label{eqxiest}
\end{equation}
(where this last equation assumes an equal number of data and random
galaxies).  We bin galaxy pairs by the absolute value of the
line-of-sight separation, i.e.\ $\pi \equiv |\pi|$.  The
``real-space'' correlation function (independent of the redshift-space
distortion) can be obtained by summing equation \ref{eqxiest} over
$\pi$.  We first define the projected correlation function
$\Xi(\sigma)$ (Davis \& Peebles 1983):
\begin{equation}
\Xi(\sigma) = 2 \sum_{\pi = 0}^{\infty} \xi_z(\sigma,\pi) \,
\Delta \pi
\label{eqxiproj}
\end{equation}
where the factor of 2 extrapolates the result to the full range
$-\infty < \pi < \infty$.  If we assume that the real-space
correlation function $\xi_r$ is well-described by a power-law
$\xi_r(r) = (r_0/r)^\gamma$, where $r_0$ is the clustering length,
$\gamma$ is the slope and $r = \sqrt{\sigma^2 + \pi^2}$, and if we
neglect the coherent infall velocities described below, we can then
derive
\begin{equation}
\xi_r(r) = \frac{\Xi(r)}{r \, C_\gamma}
\label{eqxireal}
\end{equation}
where
\begin{equation}
C_\gamma = \int_{-\infty}^\infty (1+u^2)^{-\gamma/2} du = \frac{\Gamma(\frac{1}{2}) \Gamma(\frac{\gamma-1}{2})}{\Gamma(\frac{\gamma}{2})}
\end{equation}
The difficulty with this method is that the measurement of
$\xi_z(\sigma,\pi)$ becomes noisy for large $\pi$ and therefore the
summation in equation \ref{eqxiproj} must be truncated at some $\pi =
\pi_{\rm max}$, invalidating equation \ref{eqxireal}.  We therefore
adopted the following approach (similar to the methodology of Coil et
al.\ 2008) to convert a model real-space correlation function
$\xi_r(r) = (r_0/r)^\gamma$ into a projected correlation function
which may be compared with the data.  In the linear regime, the effect
of coherent infall velocities on the correlation function can be
modelled by
\begin{equation}
\xi_z(\sigma,\pi) = \xi_0(r) P_0(\mu) + \xi_2(r) P_2(\mu) + \xi_4(r) P_4(\mu)
\end{equation}
where $P_\ell(\mu)$ are the Legendre polynomials, $\mu =
\cos{\theta}$ and $\theta$ is the angle between $r$ and $\pi$.  For a
power-law real-space correlation function,
\begin{eqnarray}
  \xi_0(r) &=& \left( 1 + \frac{2\beta}{3} + \frac{\beta^2}{5} \right) \xi_r(r) \\
\xi_2(r) &=& \left( \frac{4\beta}{3} + \frac{4\beta^2}{7} \right) \left( \frac{\gamma}{\gamma - 3} \right) \xi_r(r) \\
\xi_4(r) &=& \frac{8\beta^2}{35} \left[ \frac{\gamma(2+\gamma)}{(3-\gamma)(5-\gamma)} \right] \xi_r(r)
\end{eqnarray}
where $\beta \approx \Omega_{\rm m}(z)^{0.55}/b$ is the redshift-space
distortion parameter (Hamilton 1992; Hawkins et al.\ 2003) and $b$ is
the linear galaxy bias parameter.  We assumed $\beta = 0.6$ for this
model, consistent with our measurements (see Section
\ref{secreddist}), and for each set of trial values $(r_0,\gamma)$ we
employed the above set of equations to calculate $\xi_z(\sigma,\pi)$.
For each value of $\sigma$ we then integrated this function in the
$\pi$ direction up to $\pi = \pi_{\rm max}$ in order to compare with
the correlation function measurements.  We assumed $\pi_{\rm max} = 20
\, h^{-1}$ Mpc, and we checked that our results did not depend
sensitively on the value of $\pi_{\rm max}$.

\begin{figure}
\center
\epsfig{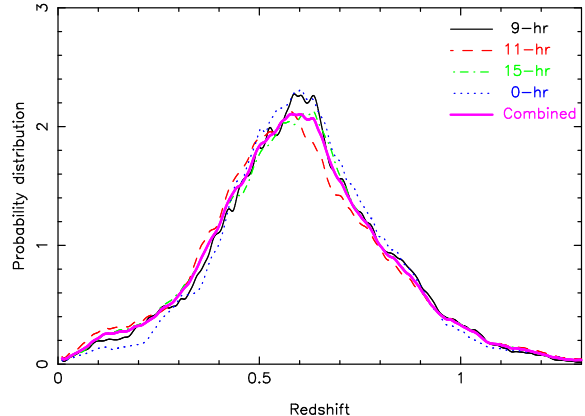}
\caption{The redshift probability distribution of WiggleZ targets with
  reliable redshifts in the four survey regions analyzed in this study
  (normalized such that $\int P(z) \, dz = 1$).  We plot cubic spline
  fits to the redshift distribution.  We also show the result for the
  combined regions as the thicker line.}
\label{fignz}
\end{figure}

We treated each of the four survey regions independently, measuring
the correlation function and corresponding error.  We then constructed
the ``combined'' correlation function by averaging the measurements in
the four regions with inverse-variance weighting.  For convenience, we
plot projected correlation functions in this paper as
$\Xi(\sigma)/(\sigma \, C_{\gamma,{\rm reduced}}) \propto
(r_0/\sigma)^\gamma$, where
\begin{equation}
C_{\gamma,{\rm reduced}} = \int_{-\sigma/\pi_{\rm max}}^{\sigma/\pi_{\rm max}} (1+u^2)^{-1/2} du
\end{equation}

\subsection{Selection function}
\label{secangsel}

We now discuss the generation of the random survey realizations that
are required for calculation of the correlation function.  This
determination of the survey ``selection function'' will be described
fully in a forthcoming paper (Blake et al.\ 2009, in preparation) and
we give a brief summary here.

\begin{figure*}
\center
\epsfig{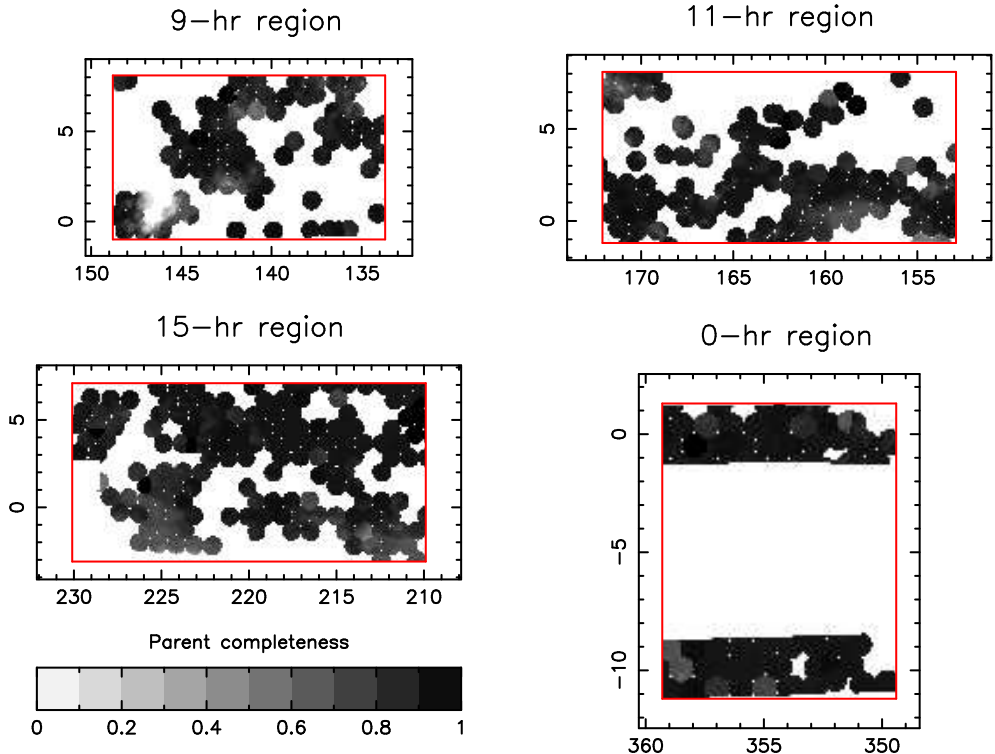}
\caption{Greyscale map illustrating the angular completeness of the
  parent catalogue of SDSS-GALEX matches for the four survey regions
  analyzed in this paper.  This parent target density varies with dust
  extinction and GALEX exposure time because the UV magnitudes of
  WiggleZ galaxies lie close to the threshold of the Medium Imaging
  Survey data.  The $x$-axis and $y$-axis of each panel are right
  ascension and declination, respectively.}
\label{figangsel}
\end{figure*}

The calculation begins with the angular selection function of the
``parent'' sample of UV-optical matches.  This function is defined
firstly by the boundaries of the GALEX fields and SDSS coverage map.
Secondly, because the UV magnitudes of our sample lie close to the
threshold of the GALEX MIS observations, there is a significant
incompleteness in the GALEX imaging that depends on the local dust
extinction and GALEX exposure time.  We used the GALEX number counts
as a function of dust and exposure time to calibrate the relation
between these quantities and the parent WiggleZ target density.  This
angular completeness function is displayed in Figure \ref{figangsel}
for the four survey regions analyzed in this paper.  We used this map
to produce a series of random realizations of the parent catalogue for
each region.

The next step is to process these random parent catalogues into random
realizations of the redshift catalogue.  The spectroscopic follow-up
of the parent catalogue comprises a network of overlapping AAOmega
pointings, with field centres optimized by the simulated annealing
algorithm and not lying on a regular grid.  The fraction of successful
redshifts in each pointing varies considerably depending on weather
conditions.  Furthermore, the redshift completeness within each
AAOmega field exhibits a significant radial variation due to
acquisition errors at the plate edges.

\begin{figure*}
\center
\epsfig{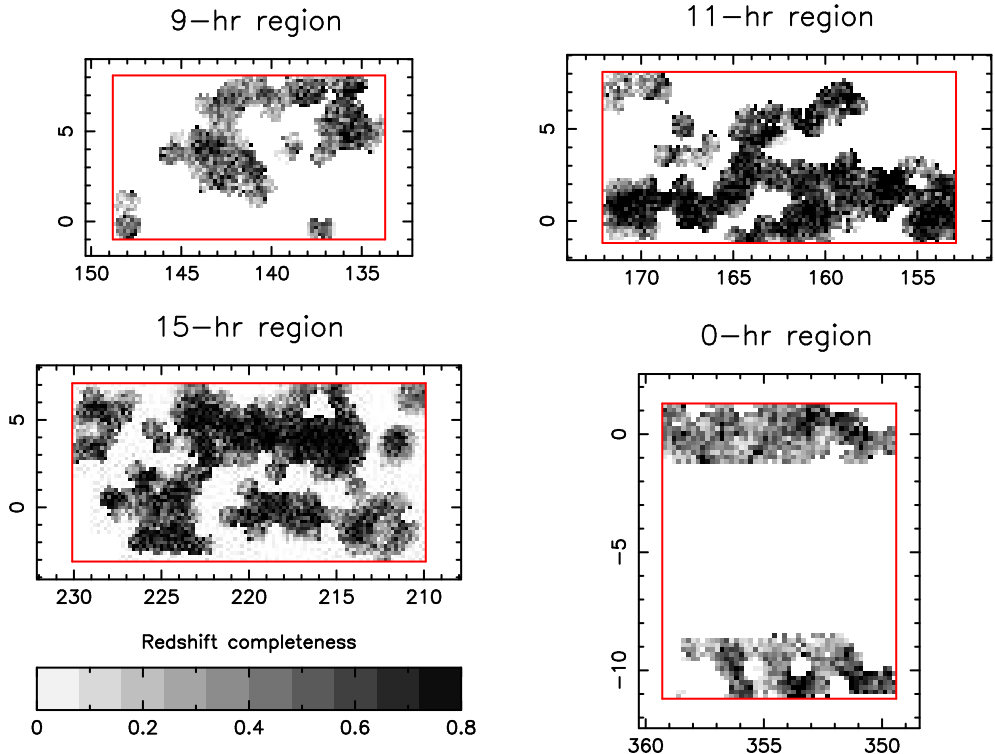}
\caption{Greyscale map illustrating the completeness of the
  spectroscopic follow-up of the WiggleZ targets shown in Figure
  \ref{figangsel} for the four survey regions analyzed in this paper.
  This Figure is generated by taking the ratio of the galaxy densities
  in the redshift and parent catalogues in small cells.  In our
  clustering analysis a more accurate approach is adopted in which the
  full AAOmega pointing sequence is applied to random realizations of
  the parent catalogue.  The $x$-axis and $y$-axis of each panel are
  right ascension and declination, respectively.}
\label{figredcomp}
\end{figure*}

In Figure \ref{figredcomp} we illustrate how the redshift completeness
varies across these survey regions by simply taking the ratio of
successful redshifts to parent galaxies in each pixel.  This is a
useful visualization, but in fact the number of unique sectors defined
by the overlapping AAOmega fields is so large that this determination
of the redshift completeness map is very noisy.  Indeed, some unique
sectors contain zero parent galaxies.

One possible approach is to smooth this completeness map over larger
areas to reduce the Poisson noise at the expense of a diminished
sensitivity to small-scale completeness variations between AAOmega
pointings.  In this analysis we use an alternative approach, which is
to apply the AAOmega pointing sequence to each of the random
realizations of the parent catalogue, and thereby create an ensemble
of random realizations of the redshift catalogue.  The AAOmega
pointing sequence is defined by the right ascension and declination of
the field centre together with the number of successful and
unsuccessful redshifts obtained for that pointing.  Within each field
centre parent galaxies are chosen randomly to create the synthetic
redshift catalogue.  It is also necessary to track the sky coverage of
the GALEX data which was contemporaneous with each AAOmega pointing.
Because the acquisition of the GALEX imaging data is ongoing with the
spectroscopic follow-up, the boundaries of the angular mask must be
modulated in step with the redshift follow-up.  In addition we impose
the radial redshift completeness variation across each AAOmega field,
measured independently for each observing run.

The redshift distribution $N(z)$ of observed galaxies varies with
position in the sky.  This is due to the magnitude prioritization
described in Section \ref{sectarsel}.  Because galaxies with fainter
$r$-band magnitudes are targeted first, the $N(z)$ will be skewed
toward higher redshifts for areas of the survey which have been
targeted by fewer AAOmega observations.  This dependence is accounted
for in our random catalogues by measuring the magnitude distribution
of targeted galaxies as a function of sky position and drawing a
random redshift from the correctly weighted $N(z)$.

\subsection{Fibre collision correction}

\begin{figure*}
\center
\epsfig{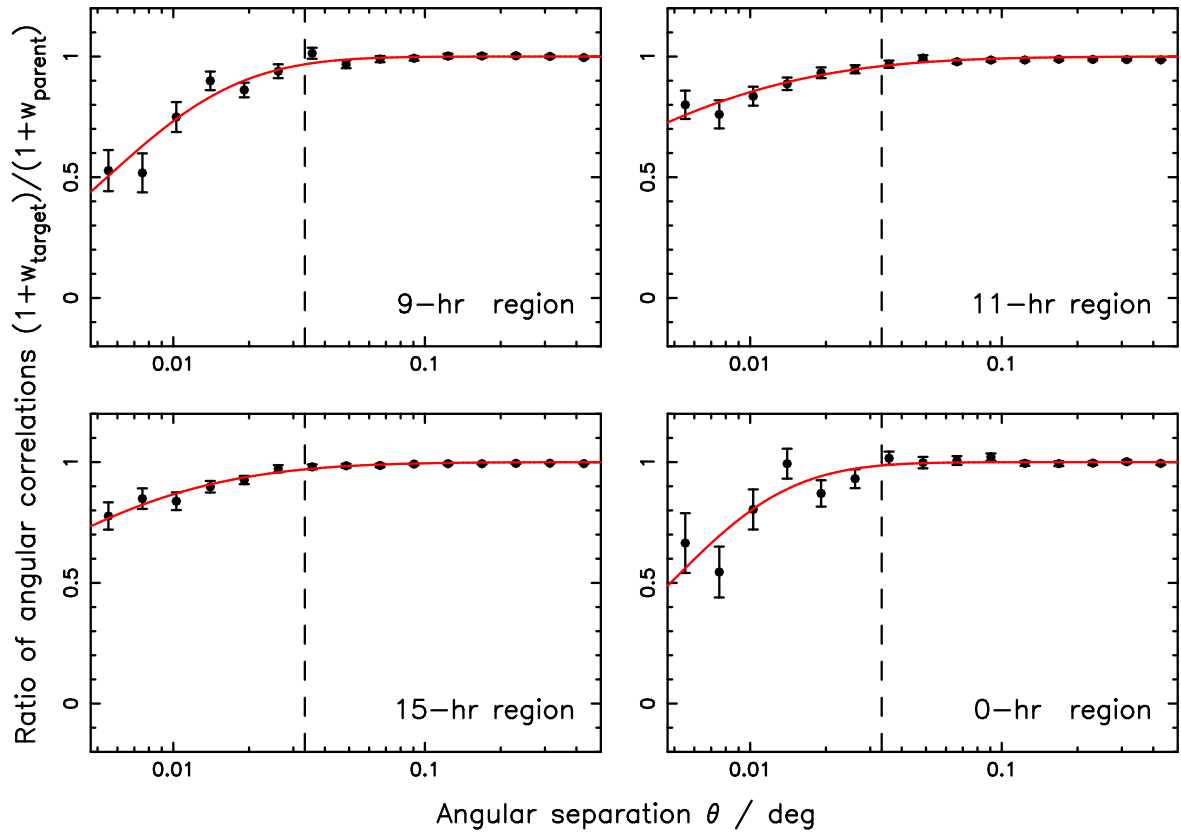}
\caption{The ratio of the angular correlation functions of the WiggleZ
  parent catalogue and targeted catalogue for the four survey regions
  analyzed in this paper.  This ratio indicates the fraction of close
  pairs surviving the restrictions of fibre collisions as a function
  of angular scale; pairs are lost for angular separations less than
  $\theta = 2$ arcmin which is indicated by the vertical dashed line.
  The solid curve indicates the best fit of the 2-parameter model $\{1
  + {\rm erf} [({\rm log_{10}}\theta-\mu)/\sigma]\}/2$.}
\label{figfibcol}
\end{figure*}

The optical fibres of the 2dF spectrograph cannot be placed closer
together than $0.5$ arcmin, and there is a diminishing probability of
observing in a single pointing both members of a close pair of parent
galaxies separated by an angular distance of less than 2 arcmin [a
  projected spatial distance of $(0.4, 0.8, 1.1, 1.4) \, h^{-1}$ Mpc
  at $z = (0.25, 0.5, 0.75, 1.0)$].  This restriction will eventually
be ameliorated by the requirement of observing each patch of sky with
3-4 AAOmega pointings to build up the number density of the redshift
catalogue.  At present, however, there is a deficit of close angular
pairs in the redshift catalogue, which artificially suppresses the
measured value of the galaxy correlation function on small scales.
The close angular pair deficit is illustrated in Figure
\ref{figfibcol} by plotting the ratio $(1+w_t)/(1+w_p)$ as a function
of angular separation $\theta$, where $w_t$ and $w_p$ are the angular
correlation functions of the targeted catalogue and the parent
catalogue, respectively.  This ratio provides the fraction of
surviving close pairs.  In order to correct the galaxy correlation
function for the missing close pairs we increased the contribution of
each galaxy pair to the data-data pair count as a function of angular
separation by a factor $(1+w_p)/(1+w_t)$ (the inverse of the quantity
plotted in Figure \ref{figfibcol}) using a 2-parameter model $\{1 +
{\rm erf} [({\rm log_{10}}\theta-\mu)/\sigma]\}/2$ fitted to the data
in Figure \ref{figfibcol}.

We note that for a survey with a redshift-dependent galaxy number
density $n(z)$, the minimum-variance correlation function measurement
for separation $s$ is achieved if galaxies are assigned a
redshift-dependent weight $w(z) = [1 + 4 \pi n(z) J_3(s)]^{-1}$ where
$J_3(s) = \int_0^s \xi(x) \, x^2 \, dx$ (Efstathiou 1988; Loveday et
al.\ 1995).  In our case the galaxy number density is sufficiently low
that $w(z) \approx 1$ and this weighting makes a negligible difference
to the results and we do not use it.

\subsection{Redshift blunder correction}

The low signal-to-noise spectra obtained by the WiggleZ survey imply
that a small but significant fraction of galaxies are assigned a
``reliable'' ($Q \ge 3$) redshift which proves to be incorrect owing
to emission-line mis-identification.  This is monitored in the survey
by allocating a small number of fibres (typically 3 to 5 out of 400
per pointing) to re-observe galaxies with existing $Q \ge 3$
redshifts.  The fraction of repeat observations producing a discrepant
redshift may be used to estimate the redshift ``blunder'' rate.

There is a significant difference in the reliability of $Q = 3$
redshifts and $Q \ge 4$ redshifts.  $Q = 3$ redshifts (which represent
a fraction $32\%$ of reliable redshifts) are typically based either on
noisy spectra or on a single emission line with no confirming spectral
features, whereas $Q \ge 4$ redshifts are based on multiple detected
emission lines.  Comparing repeat observations consisting of a $Q = 3$
redshift and a $Q \ge 4$ redshift, assuming that the latter provides
the correct redshift identification, we conclude that $\approx 17\%$
of $Q = 3$ redshifts are blunders.  This agrees with the internal
discrepancy rate amongst repeated pairs of $Q = 3$ redshifts (which is
$31\%$, which must be divided by two to obtain the blunder rate per
object).  Comparing repeat observations consisting of $Q \ge 4$
redshifts we find that only $\approx 1\%$ of these redshifts are
blunders.

\begin{figure}
\center
\epsfig{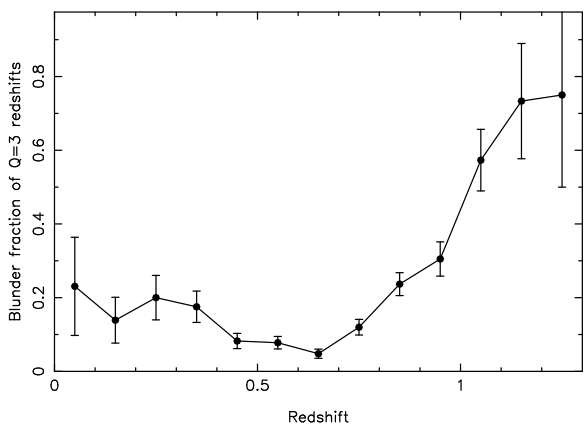}
\caption{The dependence of the redshift blunder rate of galaxies with
  $Q=3$ spectra on the (true) galaxy redshift, determined by comparing
  repeat observations consisting of pairs of spectra with $Q=3$ and $Q
  \ge 4$.  Poissonian error bars are shown.}
\label{figzblund}
\end{figure}

The blunder rate for $Q = 3$ spectra varies significantly with the
true galaxy redshift, which determines how many emission lines appear
in the observed wavelength range.  The dependence is displayed in
Figure \ref{figzblund} based on the comparison of $Q = 3$ and $Q \ge
4$ pairs of repeat observations.  The total blunder rate for all
reliable ($Q \ge 3$) redshifts is below $5\%$ for the range $z < 0.7$,
increasing to $20\%$ by $z=1$.  The redshift blunder rate does not
depend on galaxy continuum magnitude.

Redshift mis-identification reduces the measured value of the galaxy
correlation function because a fraction of true close data-data pairs
are lost as one or both of the redshifts is randomized.  If $f_{\rm
  bad}$ is the redshift blunder rate, the correction to the
correlation function is a constant factor $(1-f_{\rm bad})^{-2}$
assuming the blunder redshift is randomly distributed.  We applied
this correction to the measured correlation function to deduce the
final value:
\begin{equation}
\xi_z(\sigma,\pi)_{\rm corrected} = \xi_z(\sigma,\pi)_{\rm measured}
  \times (1 - f_{\rm bad})^{-2}
\label{eqcorrblund}
\end{equation}

When measuring the galaxy correlation function for a particular
redshift or luminosity range, we re-calculated the redshift blunder
rate for the corresponding sample in each region as explained below.
We corrected the correlation function for that region using Equation
\ref{eqcorrblund}, before combining together the correlation functions
for the different regions.  We determined the redshift blunder rate
for each region by weighting the blunder probabilities of the $N$
individual objects in that region:
\begin{equation}
f_{\rm bad} = \frac{1}{N} \left( \sum_{i=1}^N f_{{\rm bad},i} \right)
\end{equation}
For objects with $Q = 3$ we assigned the probability for each object
based on its redshift using Figure \ref{figzblund}.  For objects with
$Q \ge 4$ we assumed a blunder rate of $1\%$.

\subsection{Jack-knife re-samples}

In order to determine the error in the measured correlation function
we must characterize the statistical fluctuations in the data sample.
It is well-known that these fluctuations are not well-described by
Poisson statistics, for which the error in the pair count $DD$ in a
separation bin would be equal to $\sqrt{DD}$.  Sample variance,
geometrical edge effects and the same galaxy participating in pairs
in different separation bins cause the statistical variance of the
galaxy pair count to exceed the Poisson prediction and induce
covariances between the bins.

In this analysis we use jack-knife re-sampling to determine the
correlation function error.  In this technique the dataset is divided
into $N$ equal-area sub-regions on the sky.  The correlation function
analysis is repeated $N$ times, in each case omitting one of the
sub-regions in turn.  Labelling the different correlation function
measurements at separation $s$ as $\xi_i(s)$ from $i=1$ to $i=N$, the
covariance between separation bins $j$ and $k$ was deduced as:
\begin{eqnarray}
  C_{jk} &\equiv& \left< \xi(s_j) \, \xi(s_k) \right> - \left<
    \xi(s_j) \right> \left< \xi(s_k) \right> \\ &\approx& (N-1) \left(
    \frac{\sum_{i=1}^N \xi_i(s_j) \xi_i(s_k)}{N} - \overline{\xi(s_j)}
    \overline{\xi(s_k)} \right)
\label{eqcov}
\end{eqnarray}
where $\overline{\xi(s_j)} = \sum_{i=1}^N \xi_i(s_j)/N$.  The factor
$(N-1)$ in equation \ref{eqcov} is required because the jack-knife
re-samples are not independent, sharing a high fraction of common
sources.

We defined the jack-knife samples by splitting each survey region into
$N = 49$ sub-regions using constant boundaries of right ascension and
declination.  We tried the alternative technique of using the GALEX
tiles to define the jack-knife regions; this produced a result that
did not differ significantly.  Future analyses of the WiggleZ survey
clustering will quantify the statistical fluctuations using mock
galaxy catalogues constructed from N-body simulations.

\section{Results}
\label{secresults}

\subsection{2D correlation function}

\begin{figure}
\center
\epsfig{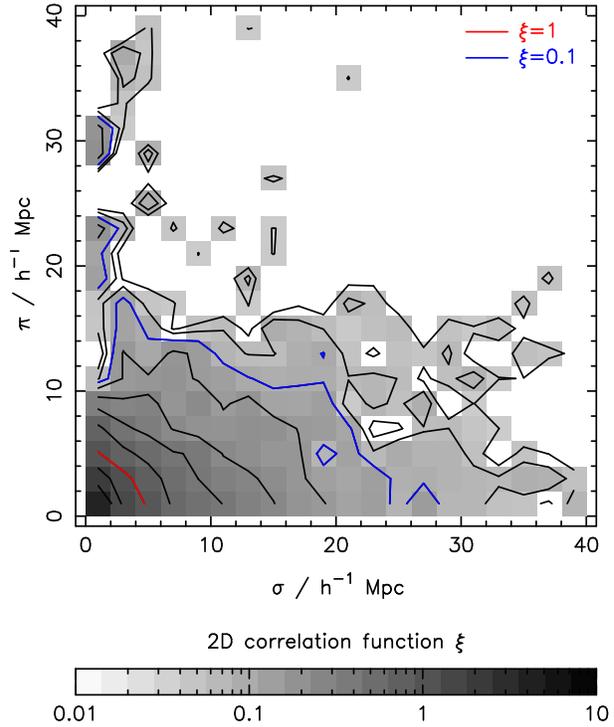}
\caption{The 2D redshift-space correlation function
  $\xi_z(\sigma,\pi)$ as a function of separation $\sigma$
  perpendicular to the line-of-sight and $\pi$ parallel to the
  line-of-sight.  The function is represented using both greyscale and
  contours.  Results for the four survey regions analyzed in this
  paper have been combined for the galaxy redshift range $0.1 < z <
  1.3$.  The non-circularity of the contours encodes the imprint of
  galaxy peculiar velocities, as discussed in the text.  The red line
  (3rd contour from the bottom left) is the $\xi_z = 1$ contour which
  lies at approximately $\sqrt{\sigma^2 + \pi^2} \approx 5 \, h^{-1}$
  Mpc; the blue line (8th contour from the bottom left) is the $\xi_z
  = 0.1$ contour.}
\label{figdist2d}
\end{figure}

Figure \ref{figdist2d} illustrates the dependence of the 2D
redshift-space correlation function $\xi_z(\sigma,\pi)$ of equation
\ref{eqxiest} on the separations $\pi$ and $\sigma$ perpendicular and
parallel to the line-of-sight for the sample of WiggleZ galaxies
spanning the full redshift range $0.1 < z < 1.3$.  We measured the
correlation function separately for the four independent survey
regions and combined the results using inverse-variance weighting.
The non-circularity of the contours of constant $\xi_z$ trace the
imprint of galaxy peculiar velocities; we use linear scales of
$\sigma$ and $\pi$ in this plot to focus on the large-scale
distortions.  In particular, for scales $> 10 \, h^{-1}$ Mpc the
increase in the value of $\xi_z$ with increasing angle to the
line-of-sight $\theta = {\rm arctan}(\sigma/\pi)$ for fixed total
separation $\sqrt{\sigma^2 + \pi^2}$ is a signature of coherent galaxy
infall and can be quantified to measure the redshift-space distortion
parameter $\beta$ (see Section \ref{secreddist}).  We also detect some
evidence for ``fingers of god'', in the form of elongation of the
contours of $\xi_z$ along the $\pi$-axis, due to the virialized
motions of galaxies in clusters.  There is some similarity here with
the results of Coil et al.\ (2008) Figure 7 for luminous blue
galaxies, except that our sample size is significantly larger.

\subsection{Clustering length of the sample}
\label{secr0}

\begin{figure}
\center
\epsfig{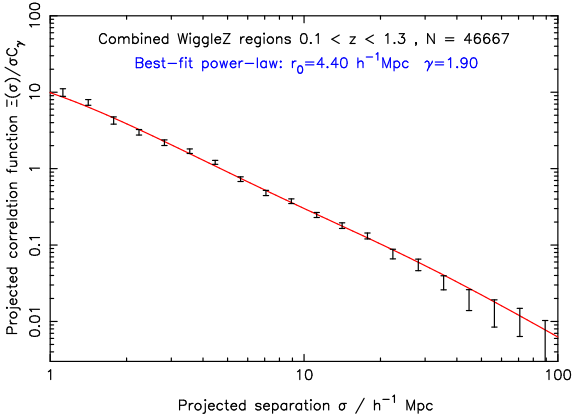}
\caption{The projected correlation function $\Xi(\sigma)/\sigma
  C_\gamma$ as a function of projected separation $\sigma$ for
  galaxies in the redshift range $0.1 < z < 1.3$, combining the
  results for the four survey regions analyzed in this paper.  The
  solid line is the best-fitting power-law for the separation range
  $1.5 < \sigma < 15 \, h^{-1}$ Mpc.  The $y$-axis is normalized by a factor
  which produces numerical results approximating
  $(r_0/\sigma)^\gamma$.}
\label{figprojall}
\end{figure}

\begin{figure}
\center
\epsfig{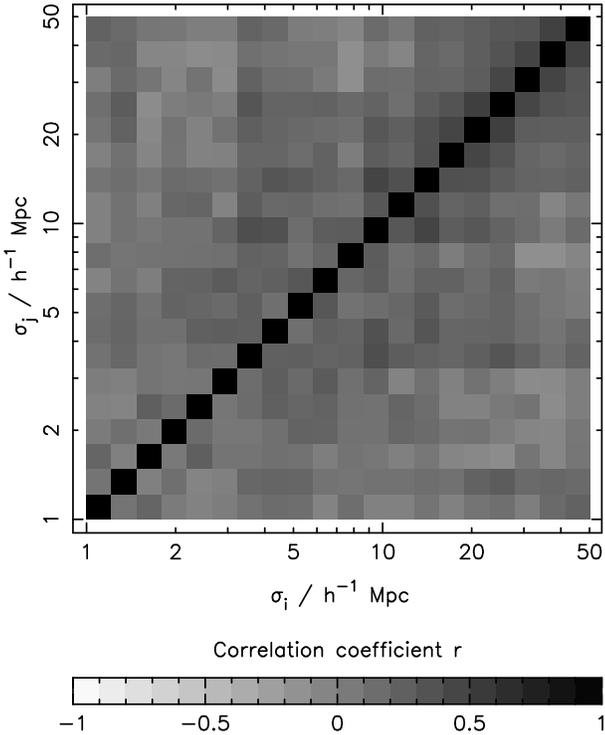}
\caption{Greyscale plot of the correlation coefficient $r$ of equation
  \ref{eqcoef}, indicating the degree of covariance between different
  separation bins for each redshift slice.}
\label{figcov}
\end{figure}

Galaxy peculiar velocities change values of $\pi$ but not $\sigma$.
The real-space clustering properties of the galaxies may therefore be
deduced by integrating $\xi_z(\sigma,\pi)$ along the $\pi$-axis, as
discussed in Section \ref{seccorrest}.  We summed the 2D correlation
function for the $0.1 < z < 1.3$ sample in 5 logarithmic bins of $\pi$
between $\pi_{\rm min} = 0.5 \, h^{-1}$ Mpc and $\pi_{\rm max} = 20 \,
h^{-1}$ Mpc.  The result is plotted in Figure \ref{figprojall} for the
projected separation range $1 < \sigma < 100 \, h^{-1}$ Mpc, with
errors obtained from the jack-knife re-sampling.  The full covariance
matrix $C$ deduced from the jack-knife re-samples is displayed in
Figure \ref{figcov} by plotting in greyscale the correlation
coefficient between two separation bins $i$ and $j$:
\begin{equation}
r(i,j) = \frac{C_{ij}}{\sqrt{C_{ii} \, C_{jj}}}
\label{eqcoef}
\end{equation}

We employed the methodology of Section \ref{seccorrest} to fit a
power-law real-space correlation function $\xi_r = (r_0/r)^\gamma$ to
the redshift-space data over the range $1.5 < \sigma < 15 \, h^{-1}$
Mpc, by minimizing the $\chi^2$ statistic using the covariance matrix:
\begin{equation}
\chi^2 = \sum_{i,j} \delta y_i \, (C^{-1})_{ij} \, \delta y_j
\end{equation}
where $\delta y_i$ is the vector of offsets between the data and the
model, and $C^{-1}$ is the inverse of the covariance matrix.  The
fitting range was motivated by our wish to estimate the clustering
length $r_0$ for which $\xi(r_0) = 1$.  A power-law provides a good
fit to the data with a best-fitting $\chi^2 = 7.1$ (for 8 degrees of
freedom).  The marginalized measurements of the power-law parameters
are $r_0 = 4.40 \pm 0.12 \, h^{-1}$ Mpc and $\gamma = 1.92 \pm 0.08$
for the $0.1 < z < 1.3$ sample.  We compare these measurements to
previous studies of UV-selected and optically-selected galaxies in
Section \ref{secprevious}.

\begin{figure*}
\center
\epsfig{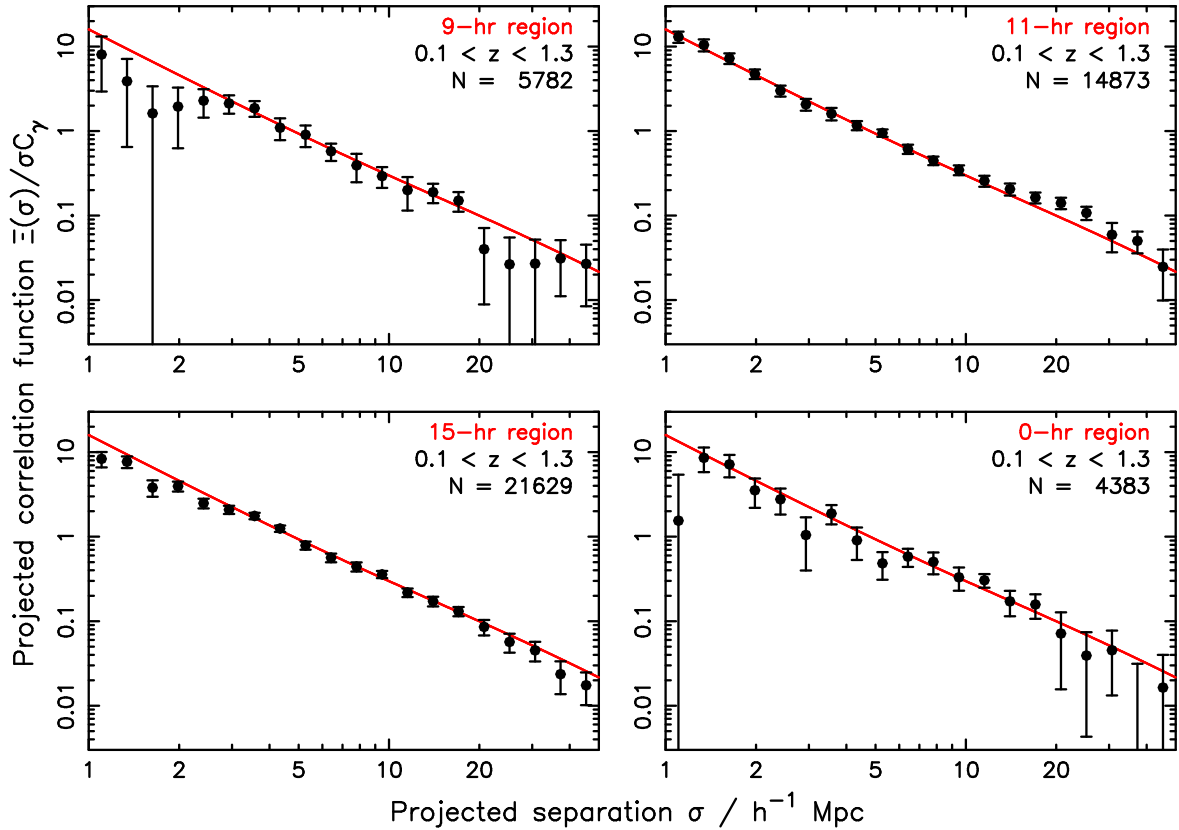}
\caption{The projected correlation function $\Xi(\sigma)/\sigma
  C_\gamma$ as a function of projected separation $\sigma$ for
  galaxies in the redshift range $0.1 < z < 1.3$, measured for the
  four survey regions analyzed in this paper.  The solid line
  indicates the best-fitting power law for the whole sample, and the
  number of redshifts $N$ used for each region is displayed.  The
  $y$-axis is normalized by a factor which produces numerical results
  approximating $(r_0/\sigma)^\gamma$.}
\label{figprojreg}
\end{figure*}

In Figure \ref{figprojreg} we plot the separate projected correlation
function measurements for each of the four survey regions analyzed in
this paper.  The four regions give consistent results.

In order to derive the bias factor of the galaxies with respect to
dark matter we generated a model non-linear matter power spectrum at
$z=0$ assuming a flat cosmological model with fiducial parameters
$\Omega_{\rm m} = 0.3$, $\Omega_{\rm b}/\Omega_{\rm m} = 0.15$, $h =
0.7$ (where $H_0 = 100 \, h$ km s$^{-1}$ Mpc$^{-1}$) and $\sigma_8 =
0.9$, using the ``{\rm CAMB}'' software package (Lewis, Challinor \&
Lasenby 2000) including corrections for non-linear growth of structure
using the fitting formula of Smith et al.\ (2003).  We used this model
power spectrum to determine the non-linear matter correlation function
$\xi_{\rm DM}$ at $z=0$.  The resulting correlation function satisfied
$\xi_{\rm DM}(r) = 1$ for $r = 4.7 \, h^{-1}$ Mpc, which we assumed as
our estimate of $r_{0,\rm{DM}}(0)$, the clustering length of dark
matter at $z=0$.  Given that the overall amplitude of the power
spectrum scales with redshift in the linear regime as $D(z)^2$, where
$D(z)$ is the linear growth factor, we can approximate:
\begin{equation}
r_{0,\rm{DM}}(z) \approx (4.7 \, h^{-1} \, {\rm Mpc}) \times
D(z)^{2/\gamma}
\end{equation}
where $\gamma \approx 1.8$.  Hence the linear bias factor $b$ of a
population of galaxies with clustering length $r_0$ can be
approximated as:
\begin{equation}
b \approx \left( \frac{r_0}{r_{0,\rm{DM}}} \right)^{\gamma/2} = \left(
\frac{r_0}{4.7 \, h^{-1} \, {\rm Mpc}} \right)^{\gamma/2} \times
D(z)^{-1}
\label{eqbias}
\end{equation}
Our measured clustering length $r_0 = 4.4 \, h^{-1}$ Mpc for a sample
at median redshift $z \approx 0.6$ is hence equivalent to a linear
bias factor $b \approx 1.3$.

\subsection{Redshift-space distortions}
\label{secreddist}

The peculiar velocities generated by large-scale coherent infall can
be parameterized by $\beta \approx \Omega_{\rm m}(z)^{0.55}/b$ where
$b$ is the linear bias parameter (Kaiser 1987).  For a flat
cosmological constant model with $\Omega_{\rm m}(0) = 0.3$,
$\Omega_{\rm m}(z=0.6) = 0.64$, and our real-space clustering
measurement $b = 1.3$ hence predicts $\beta = 0.6$ at the median
redshift of the sample.  The purpose of this Section is to demonstrate
that our data contains this self-consistent signal of peculiar
velocities (we leave detailed fits for $\beta$ to a further study).

\begin{figure}
\center
\epsfig{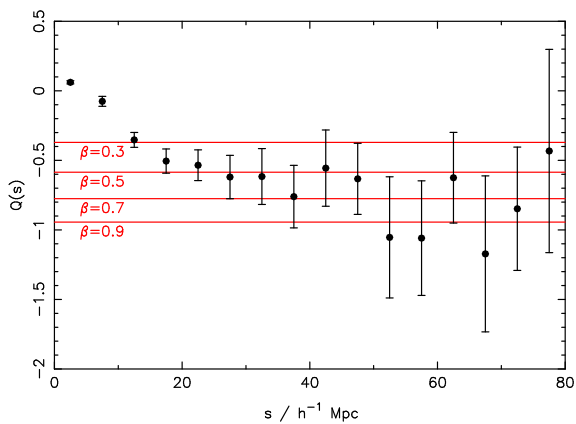}
\caption{The statistic $Q(s)$, which encodes the anisotropy in the 2D
  correlation function $\xi(\sigma,\pi)$ induced by redshift-space
  distortions.  The prediction of linear-theory on large-scales $s >
  10 \, h^{-1}$ Mpc is indicated as a function of the parameter
  $\beta$.}
\label{figdistq}
\end{figure}

We may quantify the imprint of peculiar velocities by measuring the
quadrupole moment, $Q(s)$, of the 2D correlation function (Hamilton
1992).  This statistic quantifies the anisotropy evident in Figure
\ref{figdist2d}.  If we define the correlation function moment
$\xi_\ell$ for multipole $\ell$ as:
\begin{equation}
\xi_\ell(s) = \frac{2\ell + 1}{2} \int_{-1}^{+1} \xi_z(s,\mu) \,
P_\ell(\mu) \, d\mu
\label{eqxil}
\end{equation}
we can then show that
\begin{equation}
Q(s) = \frac{\xi_2(s)}{\left[ \frac{3}{s^3} \int_0^s \xi_0(x) \, x^2
    \, dx \right] - \xi_0(s)} = \frac{ \frac{4}{3} \beta + \frac{4}{7}
  \beta }{1 + \frac{2}{3} \beta + \frac{1}{5} \beta^2}
\label{eqdistq}
\end{equation}
which is valid for large scales $s > 10 \, h^{-1}$ Mpc.  Figure
\ref{figdistq} plots the measured quantity $Q(s)$ as a function of
separation $s$, together with the prediction of equation \ref{eqdistq}
for various values of $\beta$.  In order to construct the quantity
$Q(s)$ we measured the 2D redshift-space correlation function in bins
of $s$ and $\mu$, and summed over $\mu$, weighting in accordance with
equation \ref{eqxil}.  The result is consistent with our estimate
$\beta \approx 0.6$ and constitutes a statistically-significant
detection of redshift-space distortions in our sample.

\subsection{Redshift and luminosity dependence}

Our sample of WiggleZ galaxies is large enough for us to analyze the
dependence of the clustering length $r_0$ on redshift, galaxy
luminosity and colour.  The situation is complicated by our joint
UV-optical selection and strong luminosity-redshift correlation, but
we can make some comparisons with previous studies.  We fix the
correlation function slope $\gamma = 1.8$ in this section of the
analysis.

\begin{figure}
\center
\epsfig{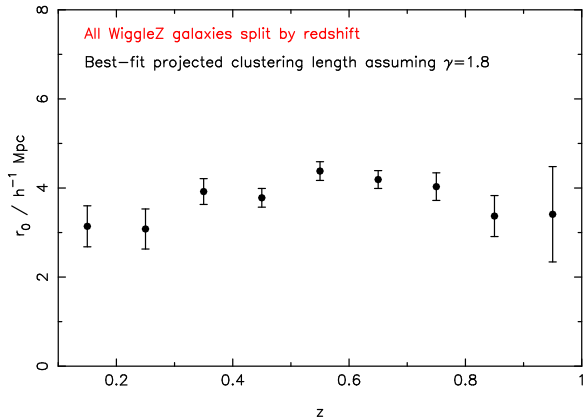}
\caption{Dependence of the best-fitting clustering length $r_0$ on
  redshift for a fit of the power-law $(r_0/r)^{1.8}$ to the
  real-space projected correlation function.}
\label{figr0vsz}
\end{figure}

The variation of the clustering length with redshift is plotted in
Figure \ref{figr0vsz}, dividing all WiggleZ galaxies in the range $0.1
< z < 1.0$ into redshift bins of width $\Delta z = 0.1$.  The
clustering length is roughly constant with redshift for the range $z >
0.3$, with a trend to a reduced clustering strength at low redshifts.
Our interpretation of the overall constancy of $r_0(z)$ is that it is
a product of two cancelling effects.  Galaxy luminosity increases with
redshift, which would tend to increase clustering length, but at
redshifts $z > 0.5$ optically red galaxies, which possess enhanced
clustering strengths, are removed from the sample by the optical
colour cuts described in Section \ref{sectarsel}.

We also analyzed the clustering in absolute magnitude and rest-frame
colour bins.  We considered the clustering as a function of rest-frame
$FUV$-band and $B$-band absolute magnitudes, which are well-matched in
wavelength (for redshift $z \approx 0.5$) to the observed-frame
$NUV$-band and $r$-band magnitudes which are used to define our target
samples.  For this initial analysis we assumed redshift-dependent
average K-corrections which we applied to all galaxies regardless of
colour.  These K-corrections were derived using the spectral energy
distribution of a Lyman Break Galaxy including an intrinsic dust
contribution $A_V = 0.14$, which produces a very good match to the
redshift-dependence of the average observed $NUV-r$ colour of the
WiggleZ targets.

\begin{figure}
\center
\epsfig{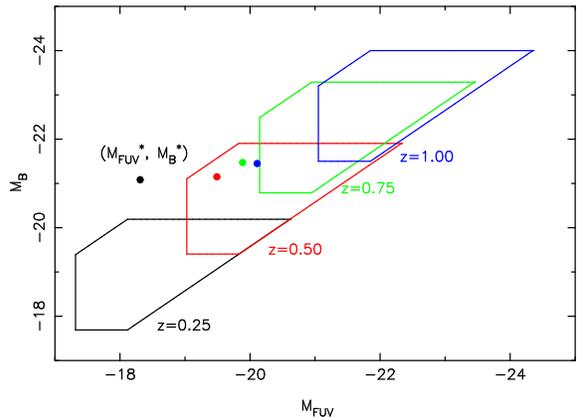}
\caption{The WiggleZ UV-optical target selection box in the space of
  $B$-band absolute magnitude $M_B$ and $FUV$-band absolute magnitude
  $M_{FUV}$ for 4 different redshifts between $z=0.25$ and $z=1$
  (moving from left to right in the Figure).  These absolute magnitude
  limits are implied by our apparent magnitude and colour selections
  $NUV < 22.8$, $20 < r < 22.5$ and $-0.5 < NUV-r < 2$.  The values of
  $M_B^*$ and $M_{FUV}^*$ at each redshift are shown for comparison
  (taken from Willmer et al.\ 2005 and Arnouts et al.\ 2005).
  Absolute magnitudes are calculated assuming $h=0.7$.}
\label{figselbox}
\end{figure}

We note that the $FUV$-band and $B$-band absolute magnitudes of our
target sample correlate strongly with redshift.  This is depicted by
Figure \ref{figselbox} which plots the target selection box in
$(M_{FUV},M_B)$ for 4 different redshifts, also indicating the
characteristic absolute magnitudes $(M_{FUV}^*,M_B^*)$ at each
redshift obtained from Arnouts et al.\ (2005) and Willmer et
al.\ (2006).  Between $z=0.25$ and $z=1$ the average value of $M_{FUV}
- M_{FUV}^*$ brightens by 2 magnitudes (becoming positive at $z
\approx 0.5$) and the average value of $M_B - M_B^*$ brightens by 4
magnitudes (becoming positive at $z \approx 0.7$).

\begin{figure}
\center
\epsfig{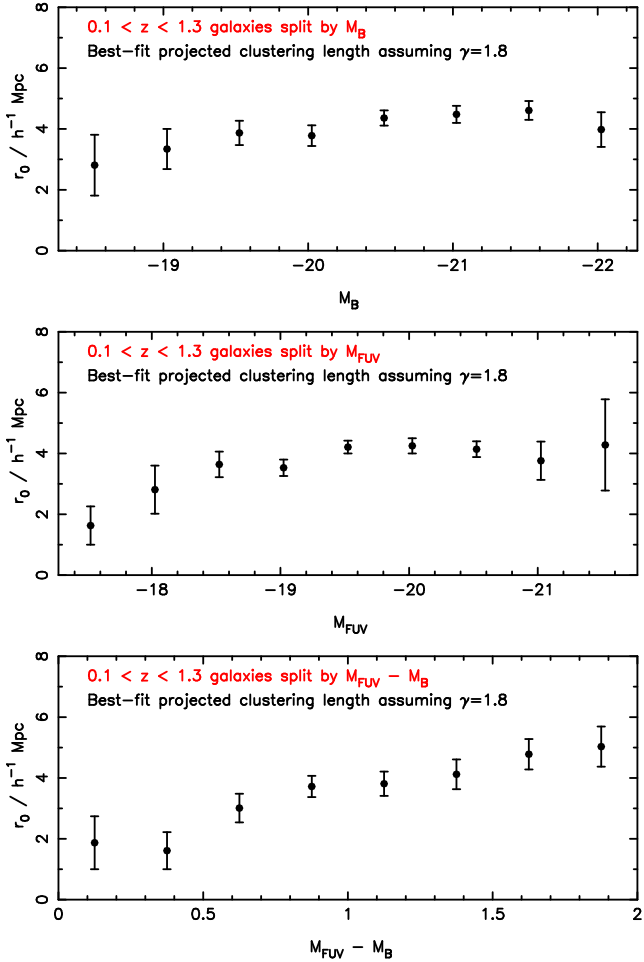}
\caption{Dependence of the best-fitting clustering length $r_0$ on
  $B$-band absolute magnitude $M_B$, $FUV$-band absolute magnitude
  $M_{FUV}$ and rest-frame colour $M_{FUV} - M_B$, for a fit of the
  power-law $(r_0/r)^{1.8}$ to the real-space projected correlation
  function.  Absolute magnitudes are calculated assuming $h=0.7$.}
\label{figr0panel}
\end{figure}

The dependence of the clustering length $r_0$ of the $0.1 < z < 1.3$
WiggleZ sample on $M_B$, $M_{FUV}$ and $M_{FUV} - M_B$ is displayed in
the panels of Figure \ref{figr0panel}.  These measurements show that
the clustering strength of the sample increases steadily with $B$-band
absolute magnitude, $FUV$-band absolute magnitude and reddening
$M_{FUV} - M_B$ colour. Sub-samples have values of $r_0$ ranging from
$2 \, h^{-1}$ Mpc to $5 \, h^{-1}$ Mpc.

\begin{figure}
\center
\epsfig{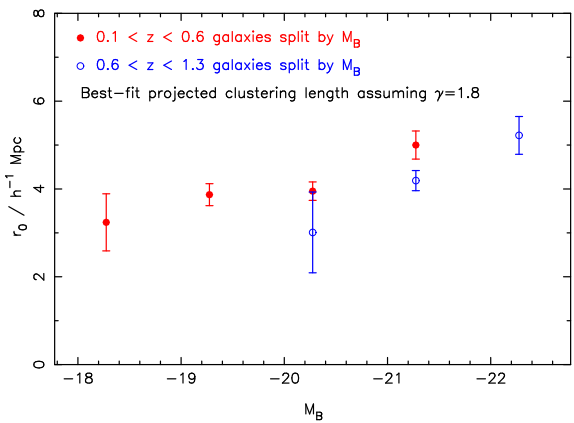}
\caption{Dependence of the best-fitting clustering length $r_0$ on
  $B$-band absolute magnitude $M_B$ for the upper and lower redshift
  ranges of our sample, for a fit of the power-law $(r_0/r)^{1.8}$ to
  the real-space projected correlation function.  Absolute magnitudes
  are calculated assuming $h=0.7$.}
\label{figr0vsmb}
\end{figure}

Figure \ref{figr0vsmb} plots the variation of $r_0$ with $M_B$ for the
low-redshift and high-redshift halves of the dataset, divided at $z =
0.6$.  This measurement confirms that at fixed $M_B$, the clustering
length of the sample drops slightly with redshift as the redder
galaxies are removed by the colour cuts.

\subsection{Comparison to previous studies}
\label{secprevious}

Coil et al.\ (2008) present clustering measurements as a function of
galaxy colour and luminosity for the DEEP2 Galaxy Redshift Survey,
which has measured redshifts for $\approx 30{,}000$ galaxies in the
range $0.7 < z < 1.5$ over an area of 3 deg$^2$.  The DEEP2 subset of
luminous blue galaxies (Coil et al.\ Table 2, line 5) has best-fitting
clustering parameters $r_0 = (4.27 \pm 0.43) \, h^{-1}$ Mpc and
$\gamma = 1.75 \pm 0.13$ at $z = 1$ (for a galaxy density $n = 6
\times 10^{-4} \, h^3$ Mpc$^{-3}$ and median absolute magnitude $M_B =
-22.1$ assuming $h=0.7$).  These results lie in good agreement with
ours.

Milliard et al.\ (2007) and Heinis et al.\ (2007) present clustering
analyses of GALEX-selected samples.  At low redshift ($z < 0.3$) the
clustering strength of the UV-selected sample is $r_0 \approx 3.5 \,
h^{-1}$ Mpc, corresponding to low-density environments, and shows no
dependence on UV luminosity (indeed, there is tentative evidence for
an anti-correlation between $r_0$ and luminosity).  These results may
naturally be compared to clustering measurements of $z \approx 3$ LBGs
also selected at rest-frame UV wavelengths (e.g.\ Giavalisco \&
Dickinson 2001; Ouchi et al.\ 2001; Arnouts et al.\ 2002; Foucaud et
al.\ 2003; Adelberger et al.\ 2005; Allen et al.\ 2005; Ouchi et
al.\ 2005; Lee et al.\ 2006; Yoshida et al.\ 2008).  These results
show a qualitatively different conclusion: LBGs are highly clustered
and concentrated in overdense regions.  Furthermore, the clustering
strength for galaxies brighter than $M_{FUV}^*$ increases with $FUV$
luminosity, reaching $r_0 \approx 15 \, h^{-1}$ Mpc for the most
luminous sub-samples.  Yoshida et al.\ (2008) demonstrate that the
behaviour of the clustering length $r_0$ depends on a combination of
UV and optical luminosities: galaxies bright in optical magnitudes are
strongly clustered irrespective of UV magnitude, whereas galaxies
faint in optical magnitude have correlation lengths increasing with UV
luminosity (see Yoshida et al.\ Fig.15).

\begin{figure}
\center
\epsfig{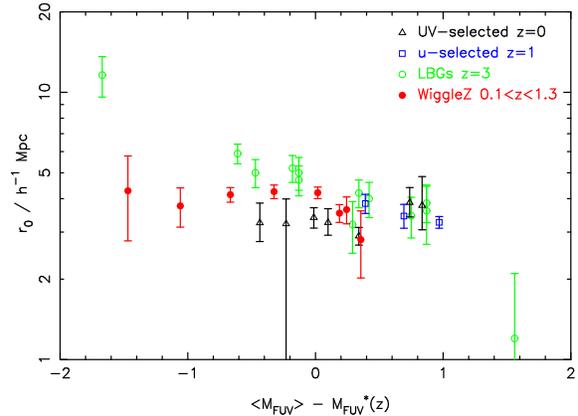}
\caption{Comparison of the clustering segregation with FUV absolute
  magnitude observed in the WiggleZ sample with the compilation of
  low-redshift and high-redshift results presented by Heinis et
  al.\ (2007).  The WiggleZ targets are more comparable to $z=3$ LBGs
  rather than $z=0$ UV-selected galaxies.  The displayed data points
  are obtained from Giavalisco \& Dickinson (2001), Arnouts et
  al.\ (2002), Foucaud et al.\ (2003), Heinis et al.\ (2004),
  Adelberger et al.\ (2005) and Heinis et al.\ (2007).  Absolute
  magnitudes are calculated assuming $h=0.7$.}
\label{figr0compare}
\end{figure}

In Figure \ref{figr0compare} we overplot the clustering measurements
of the $0.1 < z < 1.3$ WiggleZ sample as a function of $FUV$ absolute
magnitude on the compilation of low-redshift and high-redshift
clustering measurements presented by Heinis et al.\ (2007).  At low
$FUV$ absolute magnitudes $M_{FUV} - M_{FUV}^* > 0.5$ the clustering
strengths of the different UV-selected samples agree well.  This
absolute magnitude range corresponds to low redshifts $z < 0.3$ in the
WiggleZ sample (Figure \ref{figselbox}) for which we recover a
clustering length $r_0 \approx 3 \, h^{-1}$ Mpc, similar to samples of
low-redshift quiescent star-forming galaxies.  At higher $FUV$
luminosities and redshifts, the WiggleZ clustering strength is more
comparable to $z=3$ LBGs rather than $z=0$ UV-selected galaxies.  This
is expected as the $FUV-NUV$ WiggleZ selection cut becomes effective
for $z > 0.3$ and the nature of the resulting WiggleZ galaxy
population changes to merger-induced starbursts.  The WiggleZ sample
does not recover the very high values of $r_0$ present in very
luminous LBGs at $z=3$; we suggest that this may be a result of the
WiggleZ colour cuts selecting against redder galaxies.

\section{Forecasts for WiggleZ survey}
\label{secforecast}

The clustering amplitude of the WiggleZ target sample is a required
input for forecasting the accuracy with which the full 1000 deg$^2$
survey will measure the large-scale galaxy power spectrum.  The error
in the power spectrum measurement $\delta P_{\rm gal}$ at a given
redshift $z$ and Fourier wavenumber $k$ is determined by the quantity
$n \times P_{\rm gal}$, where $n(z)$ is the galaxy number density and
$P_{\rm gal}(k,z)$ is the galaxy power spectrum amplitude.  This
quantity fixes the balance between sample variance and shot noise in
the measurement error such that
\begin{equation}
\frac{\delta P_{\rm gal}}{P_{\rm gal}} = \frac{1}{\sqrt{m}} \left( 1 +
\frac{1}{n \, P_{\rm gal}} \right)
\end{equation}
where $m$ is the total number of independent Fourier modes
contributing towards the measurement (e.g.\ Feldman, Kaiser \& Peacock
1994; Tegmark 1997).  The contributions of sample variance and shot
noise are equal when $n \times P_{\rm gal} = 1$.  We model the
angle-averaged redshift-space linear galaxy power spectrum as a
function of $k$ and $z$ as:
\begin{equation}
P_{\rm gal}(k,z) = P_{\rm DM}(k,0) \left[ \frac{r_{0,{\rm
        gal}}(z)}{r_{0,{\rm DM}}(0)} \right]^\gamma \left( 1 +
\frac{2\beta}{3} + \frac{\beta^2}{5} \right)
\label{eqpkmod}
\end{equation}
where we assume $r_{0,{\rm DM}}(0) = 4.7 \, h^{-1}$ Mpc, $r_{0,{\rm
    gal}}(z) = 4.4 \, h^{-1}$ Mpc, $\gamma = 1.9$ and $\beta = 0.6$.
The second term on the right-hand-side of equation \ref{eqpkmod}
describes the boost from the galaxy linear bias factor $b$ (equation
\ref{eqbias}) using the relation $P_{\rm gal} = P_{\rm DM} b^2 D^2$.
The third term is the result of redshift-space distortions averaged
over angles.  We used the cosmological parameters as listed in Section
\ref{secr0} to produce the $z=0$ dark matter power spectrum:
$\Omega_{\rm m} = 0.3$, $\Omega_{\rm b}/\Omega_{\rm m} = 0.15$, $h =
0.7$ and $\sigma_8 = 0.9$.  In order to incorporate the fraction of
redshift blunders $f_{\rm bad}$ we reduced the effective value of the
power spectrum by a factor $(1 - f_{\rm bad})^2$ [i.e.\ increased the
value of $r_{0,{\rm DM}}(0)$ by a factor $(1 - f_{\rm
  bad})^{-2/\gamma}$].

\begin{figure}
\center
\epsfig{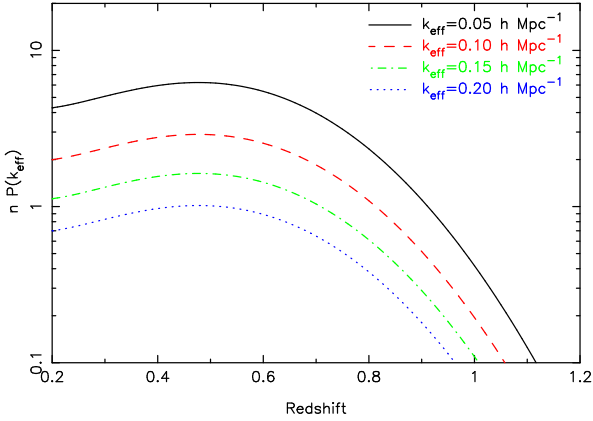}
\caption{The dependence of $n \times P_{\rm gal}$ on redshift for four
  scales $k$ representative of those important for the measurement of
  baryon acoustic oscillations.  If $n \times P_{\rm gal} = 1$, then
  the contribution of shot noise to the power spectrum error equals
  that of sample variance.}
\label{fignpvsz}
\end{figure}

Figure \ref{fignpvsz} plots the dependence of $n \times P_{\rm gal}$
on redshift for a set of different scales $0.05 < k < 0.2 \, h$
Mpc$^{-1}$ relevant for the detection of baryon acoustic oscillations,
assuming a source redshift distribution combining the survey regions
plotted in Figure \ref{fignz}.  We further assume a total target
density of 350 deg$^{-2}$ with a $70\%$ redshift completeness.  We
note that over a significant range of redshifts and scales our
large-scale power spectrum measurement will be limited by sample
variance rather than shot noise, i.e.\ $n \times P_{\rm gal} > 1$.

A useful quantity to describe the survey is the scale-dependent
``effective volume'' $V_{\rm eff}(k)$ which is defined by
\begin{equation}
V_{\rm eff}(k) = \int_0^\infty \left[ \frac{n(z) P_{\rm gal}(k,z)}{1 +
    n(z) P_{\rm gal}(k,z)} \right]^2 \frac{dV}{dz} \, dz
\end{equation}
where $dV/dz$ is the co-moving volume element.  The effective volume
represents an optimally-weighted stacking of power spectrum
measurements at different redshifts (Tegmark 1997).  For scales $k =
(0.05, 0.1, 0.15, 0.2) \, h$ Mpc$^{-1}$ we find $V_{\rm eff} = (0.65,
0.41, 0.25, 0.15) \, h^{-3}$ Gpc$^3$.  Thus the survey design will
achieve the goal of mapping $\sim 1$ Gpc$^3 = 0.34 \, h^{-3}$ Gpc$^3$.

\begin{figure}
\center
\epsfig{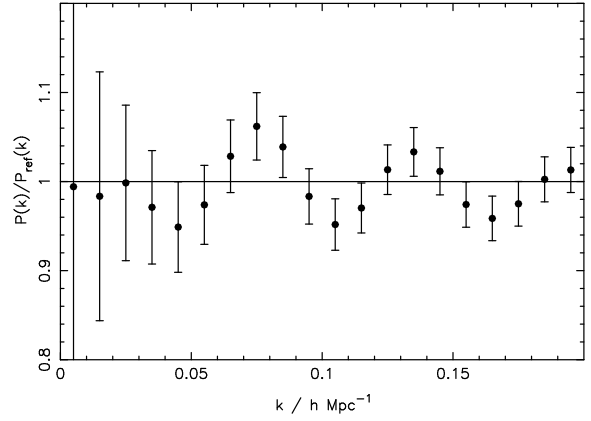}
\caption{Simulation of the errors in the final WiggleZ survey galaxy
  power spectrum.  We have divided by a smooth ``reference'' power
  spectrum to clarify the signature of baryon acoustic oscillations.}
\label{figpksim}
\end{figure}

We can use the effective survey volume to forecast the error in the
final survey power spectrum $\delta P_{\rm gal}(k)$ in a Fourier bin
of width $\Delta k$ (Tegmark 1997):
\begin{equation}
  \frac{\delta P_{\rm gal}}{P_{\rm gal}} = \frac{2 \pi}{k \sqrt{V_{\rm eff}(k) \, \Delta k}}
\label{eqpkerr}
\end{equation}
This prediction is plotted for bins of width $\Delta k = 0.01 \, h$
Mpc$^{-1}$ in Figure \ref{figpksim}, in which we divide the power
spectrum by the ``no-wiggles'' reference spectrum provided by
Eisenstein \& Hu (1998) in order to delineate clearly the baryon
acoustic oscillations.

We also generated 100 Monte Carlo realizations of the final 1000
deg$^2$ survey using the methods described in Blake \& Glazebrook
(2003) and Glazebrook \& Blake (2005).  The scatter in the power
spectrum measurements across the realizations was very close to that
predicted by equation \ref{eqpkerr}.  We used these Monte Carlo
realizations to assess the accuracy with which the full WiggleZ survey
will measure the tangential and radial standard ruler scale imprinted
by the baryon acoustic oscillations via the fitting formula described
in Blake et al.\ (2006).  Restricting ourselves to the $0.3 < z < 0.9$
subset, and first considering an ``angle-averaged'' measured power
spectrum $P(k)$, we found that the scatter in the fitted acoustic
wavescale was $2.8\%$.  Measuring instead a 2D power spectrum
$P(k_{\rm tan}, k_{\rm rad})$, where $k_{\rm tan}$ and $k_{\rm rad}$
are wavevectors measured perpendicular and parallel to the
line-of-sight, the scatters in the tangential and radial fitted
wavescales were $4.6\%$ and $7.2\%$, respectively.  This latter pair
of measurements corresponds to the accuracy of determination of the
quantities $D_A(z)/s$ and $H(z)^{-1}/s$ at an effective redshift $z
\approx 0.6$, where $D_A$ is the angular diameter distance, $H(z)$ is
the high-redshift Hubble constant, and $s$ is the sound horizon at
recombination, i.e.\ the standard ruler scale.  Dividing the survey
into redshift slices we find that the angle-averaged wavescale may be
measured with accuracy $(6.6\%, 3.7\%, 6.3\%)$ in redshift slices
$(0.25 - 0.5, 0.5 - 0.75, 0.75 - 1)$.  The angle-averaged wavescale
measures a quantity proportional to $(D_A^2 H^{-1})^{1/3}$, as
discussed by Eisenstein et al.\ (2005).

These forecasts should be considered a pessimistic lower limit on
expected performance for two reasons.  Firstly we have neglected the
cosmological information contained in the overall shape of the galaxy
power spectrum, which is divided out in the above analysis to focus on
the ``standard ruler'' aspect of the acoustic oscillations.  This
method produces robustness against systematic errors (which are
expected to affect the shape of the power spectrum but not the
oscillatory signature).  The power spectrum shape carries information
about $\Omega_{\rm m}$ and $H_0$ which further breaks the degeneracy
in cosmological distances between these two parameters and the dark
energy.  Secondly we have neglected the improvement offered by
``reconstruction'' of the density field, which sharpens the
measurement of the acoustic signature by un-doing (to first order) the
large-scale coherent galaxy motions which smooth out the acoustic
peaks (Eisenstein et al. 2007).

\begin{table*}
\center
\caption{Model WiggleZ survey parameters in one and three redshift
  bins used to forecast cosmological parameter measurements.  The bias
  factor has been multiplied by a factor $1 - f_{\rm bad}$ to
  produce an effective value allowing for the redshift blunder rate.
  The five standard ruler accuracies acc$_1$, acc$_2$, acc$_3$,
  acc$_4$, acc$_5$ are respectively the tangential and radial
  precision predicted by the Blake et al.\ (2006) fitting formula, an
  angle-averaged version of the Blake et al.\ formula, and the
  tangential and radial accuracies predicted by the Seo \& Eisenstein
  (2007) fitting formula including density reconstruction.  In the
  Blake et al.\ formula the effective bias is increased by a factor
  $\sqrt{1 + \frac{2}{3} \beta + \frac{1}{5} \beta^2} = 1.21$ to allow
  for redshift-space effects.  Further details are given in the text.}
\begin{tabular}{cccccccccc}
\hline
Redshift slice & Number density & $r_{0,{\rm gal}}$ & Bias factor & Blunder rate & acc$_1$ & acc$_2$ & acc$_3$ & acc$_4$ & acc$_5$ \\
& ($\times 10^{-4} \, h^3$ Mpc$^{-3}$) & $h^{-1}$ Mpc & $b$ & $f_{\rm bad}$ & (\%) & (\%) & (\%) & (\%) & (\%) \\
\hline
$0.3 < z < 0.9$ & 2.29 & 4.3 & 1.21 & 0.037 & 4.6 & 7.2 & 2.8 & 2.7 & 4.3 \\
\hline
$0.25 < z < 0.5$ & 3.33 & 4.0 & 1.01 & 0.038 & - & - & 6.6 & 5.5 & 8.7 \\
$0.5 < z < 0.75$ & 2.78 & 4.4 & 1.27 & 0.022 & - & - & 3.7 & 3.6 & 5.8 \\
$0.75 < z < 1$ & 0.83 & 4.4 & 1.27 & 0.127 & - & - & 6.3 & 7.9 & 10.9 \\
\hline
\end{tabular}
\label{tabmodel}
\end{table*}

We investigated improved forecasts using the methodology of Seo \&
Eisenstein (2007) which properly incorporates information from the
power spectrum shape, redshift-space distortions and density-field
reconstruction.  The predicted tangential and radial measurement
accuracies for the $0.3 < z < 0.9$ sample are $2.7\%$ and $4.3\%$,
respectively (and are correlated with a correlation coefficient $r
\approx 0.4$, further enhancing the power to constrain the
cosmological model).  We assume here that reconstruction can improve
the parameters ($\Sigma_\perp$, $\Sigma_\parallel$) defined by Seo \&
Eisenstein (2007) by a factor equal to $0.5 - 0.3 \, {\rm log}_{10}(n
\times P_{\rm gal})$ (Eisenstein, priv. comm.).  Dividing the survey
into redshift slices we find that the tangential and radial
wavescales may be measured with accuracies $(5.5\%, 8.7\%)$ for $0.25
< z < 0.5$, $(3.6\%, 5.8\%)$ for $0.5 < z < 0.75$ and $(7.9\%,
10.9\%)$ for $0.75 < z < 1$.  This information is collected in Table
\ref{tabmodel} for ease of reference.

\begin{figure}
\center
\epsfig{file=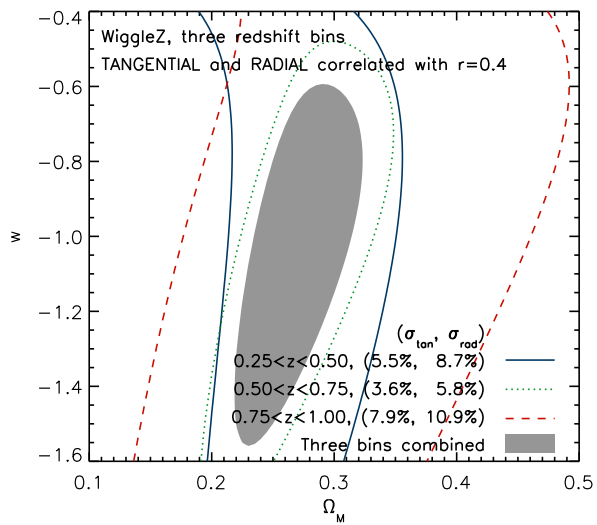,width=8.5cm,angle=0}
\caption{The forecast $68\%$ confidence ellipses for measurements of a
  constant dark energy equation-of-state $w_{\rm cons}$ and the matter
  density $\Omega_{\rm m}$ using standard ruler measurements from the
  final WiggleZ survey in combination with a CMB prior on the acoustic
  scale $\ell_A$.  Results are shown for 3 redshift bins (the
  different contours) and for the combination of the redshift bins
  (the shaded area).}
\label{figomw1}
\end{figure}

We used this last set of forecasts with reconstruction in 3 redshift
bins to determine the expected accuracy of measurement of a constant
equation-of-state $w_{\rm cons}$ of dark energy (assuming the
measurements of $D_A$ and $H^{-1}$ are correlated with coefficient $r
= 0.4$).  Confidence ellipses are displayed in Figure \ref{figomw1} in
the space of $w_{\rm cons}$ and the matter density $\Omega_{\rm m}$
for a flat cosmology with fiducial model $w_{\rm cons} = -1$ and
$\Omega_{\rm m} = 0.27$.  Results are shown for each redshift bin
separately and for the combination of all 3 bins.  In order to
generate this Figure we have used the 5-year-WMAP measurement of the
CMB acoustic scale $\ell_A = 302.1 \pm 0.9$ (Komatsu et al.\ 2009) in
order to cancel the dependence of the baryon oscillation measurement
on the sound horizon at recombination.  In Figure \ref{figomw1} we have
not included any further CMB information or other external datasets.
The marginalized errors are $\sigma(w_{\rm cons}) = 0.31$ and
$\sigma(\Omega_{\rm m}) = 0.03$.

\begin{figure}
\center
\epsfig{file=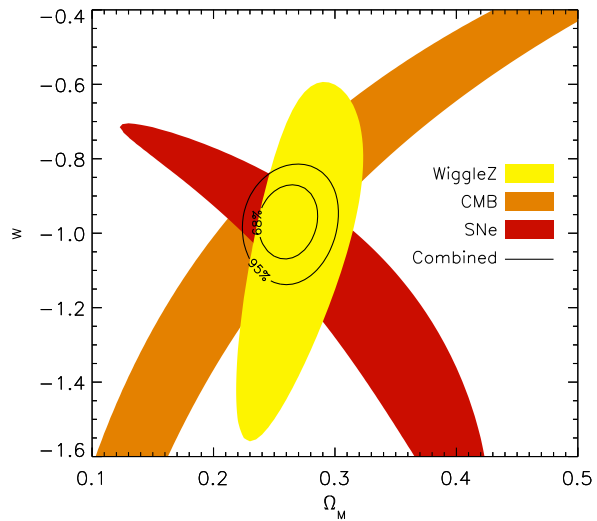,width=8.5cm,angle=0}
\caption{The forecast $68\%$ confidence ellipse for measurement of
  $(\Omega_{\rm m},w_{\rm cons})$ from the WiggleZ survey plus CMB
  acoustic scale (the yellow ellipse), compared with existing
  measurements from the CMB shift parameter (the orange ellipse) and
  latest supernovae (the red ellipse).  The $68\%$ and $95\%$
  confidence regions for the combination of all the datasets is
  displayed as the central contours.}
\label{figomw2}
\end{figure}

In Figure \ref{figomw2} we add in information from the 5-year-WMAP
measurement of the CMB shift parameter $R = 1.71 \pm 0.02$ (Komatsu et
al.\ 2009), including the correlation between $R$ and $\ell_A$,
together with the latest supernovae data from the Essence, SNLS and
HST observations (see Wood-Vasey et al.\ 2007, Astier et al.\ 2006,
Riess et al.\ 2007, Davis et al.\ 2007).  The marginalized errors in
the cosmological model from the full combination of datasets are
$\sigma(w_{\rm cons}) = 0.07$ and $\sigma(\Omega_{\rm m}) = 0.02$.
The forecast performance of the WiggleZ survey exceeds that of the
current CMB and supernovae data, but the different measurements are
also complementary, breaking degeneracies in the $(\Omega_{\rm
  m},w_{\rm cons})$-plane through independent techniques.
Disagreement between any pair of datasets would produce the
possibility of discovering non-standard physics (if it exists) or
systematic measurement errors.  The final accuracy of $w_{\rm cons}$
constitutes a robust and precise test of the dark energy model.

\section{Conclusions}
\label{secconc}

We have measured the small-scale clustering amplitude of high-redshift
bright emission-line galaxies using the first $20\%$ of spectra
from the AAT WiggleZ Dark Energy Survey ($\approx 47{,}000$ galaxies
in the redshift range $0.1 < z < 1.3$).  We have successfully
developed a methodology to generate random realizations of the survey
incorporating the currently sparse selection function and redshift
incompleteness.  We find that:
\begin{itemize}
\item The WiggleZ galaxy sample in the redshift range $0.1 < z < 1.3$
  possesses a clustering length $r_0 = 4.40 \pm 0.12 \, h^{-1}$ Mpc
  and slope $\gamma = 1.92 \pm 0.08$.  This clustering amplitude
  significantly exceeds that of UV-selected samples at $z \approx 0$
  and agrees well with that of the most luminous blue galaxies
  observed by the DEEP2 galaxy redshift survey at $z \approx 1$.  The
  clustering amplitude of WiggleZ targets is comparable to that of
  Lyman Break Galaxies at a similar UV luminosity.
\item The clustering length of the WiggleZ targets is approximately
  constant with redshift for the range $z > 0.3$.  The value of $r_0$
  increases with $B$-band luminosity, $FUV$-band luminosity, and
  reddening rest-frame colour.
\item The redshift-space distortion signature of coherent galaxy
  motions is detected and its amplitude ($\beta \approx 0.6$) is
  consistent with that predicted from the galaxy bias.  We detect some
  evidence for ``fingers of god'' due to the virialized motions of
  galaxies in clusters.
\end{itemize}
Using these results, we forecast the performance of the full 1000
deg$^2$ WiggleZ survey in the measurement of the galaxy power spectrum
and cosmological model.  We find that:
\begin{itemize}
\item The survey design is well-tuned to the ``optimal'' mean galaxy
  number density $n \sim P_{\rm gal}^{-1}$, where $P_{\rm gal}$ is the
  amplitude of the galaxy power spectrum on the scales of importance
  for baryon oscillations.
\item The survey will delineate the baryon acoustic oscillations in
  the large-scale clustering pattern in three independent redshift
  slices, providing measurements of the cosmic distance and expansion
  rate in each redshift slice with accuracies of $\approx 5\%$.
\item The resulting measurement of a constant dark energy
  equation-of-state parameter $w_{\rm cons}$ from the WiggleZ survey,
  calibrating the standard ruler using the CMB measurement of the
  acoustic scale, has a higher precision than provided by current
  supernovae datasets.  These independent dark energy probes lie in a
  highly complementary direction in the parameter space of $w_{\rm
    cons}$ and $\Omega_{\rm m}$.  The full combination of WiggleZ,
  supernovae and CMB datasets provides a measurement of the equation
  of state with accuracy $\Delta w_{\rm cons} = 0.07$, constituting a
  robust and precise test of the dark energy model incorporating
  cross-checking of systematic errors between different probes.
\end{itemize}
The final survey will enable a wide range of scientific investigations
into the cosmological model and galaxy evolution.

\section*{Acknowledgments}

We acknowledge financial support from the Australian Research Council
through Discovery Project grants funding the positions of SB, MP, GP
and TD.  We also thank the University of Queensland for supporting the
PhD scholarship of RJ.  We acknowledge the efforts of Nick Jones and
David Barnes in creating the online WiggleZ database, and Emily
Wisnioski for incorporating Principal Component Analysis sky
subtraction into the data reduction pipeline.  We thank Karl Forster
for his assistance in scheduling our GALEX observations and Sebastien
Heinis, Ted Wyder and Mark Seibert for invaluable GALEX support and
discussions.  We acknowledge correlation function modelling performed
by Carlos Contreras and Ben Jelliffe which revealed a mistake in the
submitted version of this paper.

GALEX (the Galaxy Evolution Explorer) is a NASA Small Explorer,
launched in April 2003.  We gratefully acknowledge NASA's support for
construction, operation and science analysis for the GALEX mission,
developed in co-operation with the Centre National d'Etudes Spatiales
of France and the Korean Ministry of Science and Technology.

Finally, the WiggleZ survey would not be possible without the
dedicated work of the staff of the Anglo-Australian Observatory in the
development and support of the AAOmega spectrograph, and the running
of the AAT.

\end{document}